\documentclass[pra,twocolumn,aps,amssymb,showpacs,superscriptaddress]{revtex4-1}

\usepackage{epsfig}
\usepackage{graphicx}
\usepackage{amsmath}
\usepackage{amssymb}
\usepackage{enumerate}
\usepackage{hyperref}

\usepackage{color}
\definecolor{rosso}{rgb}{1,0,0}
\definecolor{verde}{rgb}{0,1,0}
\definecolor{blue}{rgb}{0,0,1}
\definecolor{amber}{rgb}{1.0, 0.75, 0.0}
\definecolor{amber(sae/ece)}{rgb}{1.0, 0.49, 0.0}
\definecolor{verdescuro}{rgb}{0,0.5,0.5}
\definecolor{rossoscuro}{rgb}{0.7,0.3,0}
\definecolor{bluscuro}{rgb}{0.3,0,0.7}
\definecolor{magenta}{rgb}{1,0,1}

\begin{document}

\title{Fermi gas throughout the BCS-BEC crossover: A comparative study of \\ $t$-matrix approaches with various degrees of self-consistency}

\author{M. Pini}
\email{michele.pini@unicam.it}
\affiliation{School of Science and Technology, Physics Division \\ Universit\`{a} di Camerino, 62032 Camerino (MC), Italy}
\author{P. Pieri}
\email{pierbiagio.pieri@unicam.it}
\affiliation{School of Science and Technology, Physics Division \\ Universit\`{a} di Camerino, 62032 Camerino (MC), Italy}
\affiliation{INFN, Sezione di Perugia, 06123 Perugia (PG), Italy}
\author{G. Calvanese Strinati}
\email{giancarlo.strinati@unicam.it}
\affiliation{School of Science and Technology, Physics Division \\ Universit\`{a} di Camerino, 62032 Camerino (MC), Italy}
\affiliation{INFN, Sezione di Perugia, 06123 Perugia (PG), Italy}
\affiliation{CNR-INO, Istituto Nazionale di Ottica, Sede di Firenze, 50125 Firenze (FI), Italy}


\begin{abstract}

The diagrammatic $t$-matrix approximation has often been adopted to describe a dilute Fermi gas. 
This approximation, originally considered by Galitskii for a repulsive inter-particle interaction \cite{Galitskii-1958}, has later been widely utilized for an attractive Fermi gas to describe the BCS-BEC crossover from strongly overlapping Cooper pairs in weak coupling to non-overlapping composite bosons in strong coupling. 
Several variants of the $t$-matrix approximation have been considered in the literature, which are distinguished by the degree of self-consistency allowed in the building blocks of the diagrammatic structure. 
Here, we perform a systematic and comparative study of \emph{all} possible variants on the degree of self-consistency for the $t$-matrix approximation in an attractive Fermi gas, which enables us to confront their outcomes for thermodynamic and dynamical quantities on the same footing in an unbiased way. 
For definiteness, only the normal phase above the superfluid critical temperature is considered. 
The dispute that can be raised in this context, about the adequateness of introducing progressive degrees of self-consistency over and above the non-self-consistent $t$-matrix approximation for an attractive Fermi gas, parallels the recent interest in the literature on assessing the importance of various degrees of self-consistency in the context of semiconductors and insulators.
\end{abstract} 

\maketitle

\section{Introduction} 
\label{sec:introduction}
\vspace{-0.3cm}

The method of functional derivatives provides a general framework to deal with quantum many-particle systems in a non-perturbative fashion. 
This method, that was originally introduced in condensed matter by Martin and Schwinger \cite{MS-1959} and later adopted by Hedin \cite{Hedin-1965} (see also Ref.~\cite{Strinati-RNC-1988}), starts from the exact equations for the single- and two-particle Green's functions (namely, the Dyson and Bethe-Salpeter equations, respectively) and introduces approximations only at the level of the kernels of these integral equations.

A systematic method for selecting non-perturbative approximations which satisfy the conservation laws (the so-called ``conserving approximations'') has been formulated by Baym and Kadanoff \cite{BK-1961}.
This method intimately relates the kernel of the Dyson equation (namely, the single-particle self-energy) with the kernel of the Bethe-Salpeter equation. 
In both cases, these kernels are expressed as functionals of the single-particle propagator that solves the Dyson equation. 
In this context, the need for a ``$\Phi$-derivable'' choice of the single-particle self-energy and for the \emph{self-consistent} solution of both equations has been emphasized \cite{Baym-1962}.

Although the use of a fully self-consistent \emph{and} conserving approximation appears mandatory when dealing with physical problems that involve transport and localization, in other circumstances non-self-consistent or partly self-consistent approximations may provide physically more sensible results with respect to the self-consistent one. 
An example is provided by the fluctuation exchange (FLEX) approximation introduced for the repulsive Hubbard model (even at half-filling), whereby both the self-consistent and the non-self-consistent versions have been investigated \cite{Scalapino-1989,Kotliar-2005}. 
More recently, interest in comparing the results of the self-consistent vs non-self-consistent approaches has arose also in the context of the GW approximation for semiconductors and insulators. 
In this case, the non-self-consistent calculations turn out to better compare with the experimental values of the band gaps with respect to the self-consistent calculations \cite{Kresse-2018}. 
Similar conclusions have further been drawn in the context of a simpler model \cite{Berger-2018}.

In the context of a (dilute) Fermi gas with an attractive inter-particle interaction, the $t$-matrix approximation appears as a natural candidate to describe the system while it evolves throughout the BCS-BEC crossover. 
The first pioneering approach in this respect goes back to the work by Nozi\`{e}res and Schmitt-Rink (NSR) \cite{NSR-1985}, where a simplified version of the non-self-consistent $t$-matrix
approximation proved sufficient to highlight the main features of the crossover physics in the normal phase above the superfluid critical temperature.
Later on, the NSR approach was extended, either to improve on the treatment of the non-self-consistent $t$-matrix approximation \cite{PPSC-2002}, or to include various degrees of self-consistency within this approximation, ranging from partial \cite{Levin-1997,Ohashi-2012,Micnas-2014} to full \cite{Haussmann-1993,Haussmann-1994} self-consistency.
As expected, depending on the degree of self-consistency different numerical results were obtained for various physical quantities, ranging from thermodynamic to dynamic.
However, direct comparison among the results obtained with various degrees of self-consistency has been hindered by the (sometimes even drastic) numerical approximations that were introduced in the calculations on top of a specific choice about the degree of self-consistency.
For these reasons, it appears that a systematic and direct comparison of the results obtained by adopting various degrees of self-consistency on the $t$-matrix approximation is still lacking, especially when this comparison would be made in an unbiased way by retaining the same level of numerical accuracy for all different approaches.

Primary purpose of this paper is to fill this gap, by undertaking the above systematic study on \emph{all the five} degrees of self-consistency that have been considered thus far in the literature within the $t$-matrix approximation in the normal phase above the superfluid critical temperature \cite{PPSC-2002,Levin-1997,Ohashi-2012,Micnas-2014,Haussmann-1993,Haussmann-1994}.
Although the results of the present study may not lead to definite conclusions, about which one of the above five approximations could account best for the thermodynamic \emph{and} dynamical properties of a (dilute) Fermi gas undergoing the BCS-BEC crossover, it appears nevertheless interesting and relevant (if not timely) to discover how these properties get modified when passing from one to the other of these five approaches.
In addition, the high precision of the numerical calculations that we have implemented has enabled us to accurately check how the two distinct (BCS) and (BEC) limits of the crossover are recovered by the alternative $t$-matrix approaches.
Specifically, in the BEC limit the residual interaction among composite bosons extracted from our numerical calculations turns out to be in excellent agreement with the analytic estimates that we also provide
(which correct a previous analytic estimate obtained in Refs.~\cite{Haussmann-1993,Haussmann-1994} within the fully self-consistent $t$-matrix approach).
In the BCS limit, on the other hand, we have found that a partially self-consistent $t$-matrix approach, developed in Ref.~\cite{Levin-1997} and often utilized in the literature, breaks down when one avoids using the set of approximations that normally accompany its implementation.

Finally, it should be recalled that, in order to get a refined agreement with Quantum Monte Carlo and experimental data available for a Fermi gas in the unitary region of the crossover intermediate between the BCS and BEC regimes, it may in any case be required to go \emph{beyond} the $t$-matrix approximation \cite{Pisani-2018-I} and include on top of it also a class of vertex corrections associated with the Gorkov-Melik-Barkhudarov (GMB) contribution \cite{Gorkov-1961}.

The plan of the paper is as follows.
Section~\ref{sec:t-matrices} sets up the theoretical framework and specifies in details how the alternative $t$-matrix approaches with various degrees of self-consistency need to be handled.
Sections~\ref{sec:numeric-termodynamic} and \ref{sec:numeric-dynamic} report on the numerical results obtained within the above alternative $t$-matrix approaches, for the thermodynamic and dynamical quantities of interest, respectively.
Sections~\ref{sec:conclusions} gives our conclusions.
Appendix~\ref{sec:appendix-A} gives a detailed account about the numerical procedures we have adopted to achieve (partial or full) self-consistency within the various $t$-matrix approaches, for all relevant sub-units of the many-body structure.
Appendix~\ref{sec:appendix-B} discusses the optimization procedure that was found necessary to achieve the required convergence toward self-consistency, within some of the above alternative $t$-matrix approaches.
Finally, Appendix~\ref{sec:appendix-C} compares the results of two partially self-consistent $t$-matrix approaches with that of their approximate treatments usually utilized in the literature.

\section{Alternative $t$-matrix approaches} 
\label{sec:t-matrices}

In this Section, we set up the theoretical framework for the alternative $t$-matrix approaches that can be used to describe a Fermi gas with an \emph{attractive contact interaction} throughout the BCS-BEC crossover, in the normal phase at a temperature $T$ above the superfluid critical temperature $T_{c}$.
Only a balanced situation, with equal number of spin up and spin down fermions, will be considered in this paper.

\vspace{0.05cm}
\begin{center}
{\bf A. Basic equations}
\end{center}

Within this framework, the basic expressions read (in the following, we set the Planck constant $\hbar$ and Boltzmann constant $k_{B}$ equal to unity):
\begin{eqnarray}
G(k) &=& \Big(G_0(k)^{-1} -\Sigma(k)\Big)^{-1}
\label{equation-G} \\
\Sigma(k) & = & - \int \!\! \frac{d\mathbf{Q}}{(2\pi)^3} T \sum_\nu \Gamma(Q) \, G^{(\mathrm{c})}(Q-k)
\label{equation-Sigma_k} \\
\Gamma(Q) & = & - \bigg(\frac{m}{4\pi a_F} + R_{\mathrm{pp}}(Q) \bigg)^{-1}
\label{equation-Gamma_Q} \\
R_{\mathrm{pp}}(Q) & = & \!\! \int \!\!\!\frac{d\mathbf{k}}{(2\pi)^3} \Big( \! T \sum_n G^{(\mathrm{a})} (k) G^{(\mathrm{b})} (Q-k) - \frac{m}{\mathbf{k}^2} \! \Big) .
\label{equation-Rpp_Q}
\end{eqnarray}
\begin{figure}[t]
\begin{center}
\includegraphics[width=7.5cm,angle=0]{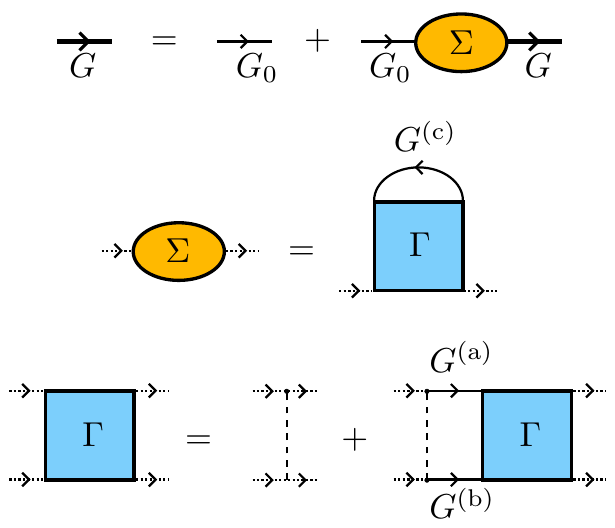}
\caption{(Color online) Diagrammatic representation of the $t$-matrix approximation. 
                                    Thick and thin lines represent the single-particle propagator $G$ and its non-interacting counterpart $G_{0}$, respectively, while broken lines stand for the 
                                    inter-particle interaction.
                                    The colored ellipse corresponds to the self-energy $\Sigma$ and the colored square to the particle-particle propagator $\Gamma$, where fermion lines connected by interaction
                                    lines are meant to have opposite spins.
                                    The positions where the three types of propagators $G^{(\mathrm{a})}$, $G^{(\mathrm{b})}$, and $G^{(\mathrm{c})}$ occur in the diagrams are also indicated.}
\label{Figure-1}
\end{center}
\end{figure} 
\noindent
Here, $G$ is the single-particle fermionic propagator and $G_{0}$ its non-interacting counterpart, $\Sigma$ is the self-energy, $\Gamma$ is the particle-particle propagator, and 
$R_{\mathrm{pp}}$ is the (regularized) particle-particle bubble.
These quantities are drawn pictorially in Fig.~\ref{Figure-1}.
In addition, $k=(\mathbf{k},\omega_{n})$ is a fermionic four-vector with wave vector $\mathbf{k}$ and fermionic Matsubara frequency $\omega_n=(2n+1)\pi \, T$ ($n$ integer),
and $Q=(\mathbf{Q},\Omega_{\nu})$ a bosonic four-vector with wave vector $\mathbf{Q}$ and bosonic Matsubara frequency $\Omega_\nu=2\pi\nu \, T$ ($\nu$ integer).
Finally, $m$ is the fermionic mass and $a_{F}$ the scattering length of the associated two-fermion problem.

Note that, in the above equations, the single-particle propagators have been distinguished as $G^{(\mathrm{a})}$, $G^{(\mathrm{b})}$, and $G^{(\mathrm{c})}$, meaning that each of them can be either the dressed $G$ or the bare $G_{0}$, in such a way that different $t$-matrix approaches can be realized by selecting different combinations of these functions.
Specifically, the short-hand notation $(G^{(\mathrm{a})}G^{(\mathrm{b})})G^{(\mathrm{c})}$ will be used to identify a given $t$-matrix approach, where the propagators $G^{(\mathrm{a})}$ and $G^{(\mathrm{b})}$ within the parentheses correspond to those entering
the particle-particle bubble (\ref{equation-Rpp_Q}), while the external propagator $G^{(\mathrm{c})}$ enters the self-energy (\ref{equation-Sigma_k}).
In the following, we shall consider the five combinations of $G^{(\mathrm{a})}$, $G^{(\mathrm{b})}$, and $G^{(\mathrm{c})}$ reported in Table~\ref{Table-I}, with the corresponding references where the various approaches have been discussed for a Fermi gas with an attractive contact interaction.

\begin{table}[h]
\begin{tabular}{ ccc } 
 \hline
 \hline
 $(G^{(\mathrm{a})} G^{(\mathrm{b})})G^{(\mathrm{c})}$ & Reference \\ 
 \hline
 $(G_{0} \, G_{0}) \, G_{0}$ \,  &   Ref.~\cite{PPSC-2002}             \\  
 $(G_{0} \, G_{0}) \, G$ \,        &    Ref.~\cite{Ohashi-2012}           \\   
 $(G \, G_{0}) \, G_{0}$ \,        &    Ref.~\cite{Levin-1997}              \\ 
 $(G \, G) \, G_{0}$ \,              &     Ref.~\cite{Micnas-2014}          \\ 
 $(G \, G) \, G$ \,                     &    Refs.~\cite{Haussmann-1993,Haussmann-1994}   \\
 \hline
 \hline
\end{tabular}
\caption{Short names adopted for the various $t$-matrix approaches, according to the conventions introduced in the expressions (\ref{equation-Rpp_Q}) and (\ref{equation-Sigma_k}), with the corresponding key references to these approaches.}
\label{Table-I}
\end{table}

\begin{figure}[t]
\begin{center}
\includegraphics[width=9cm,angle=0]{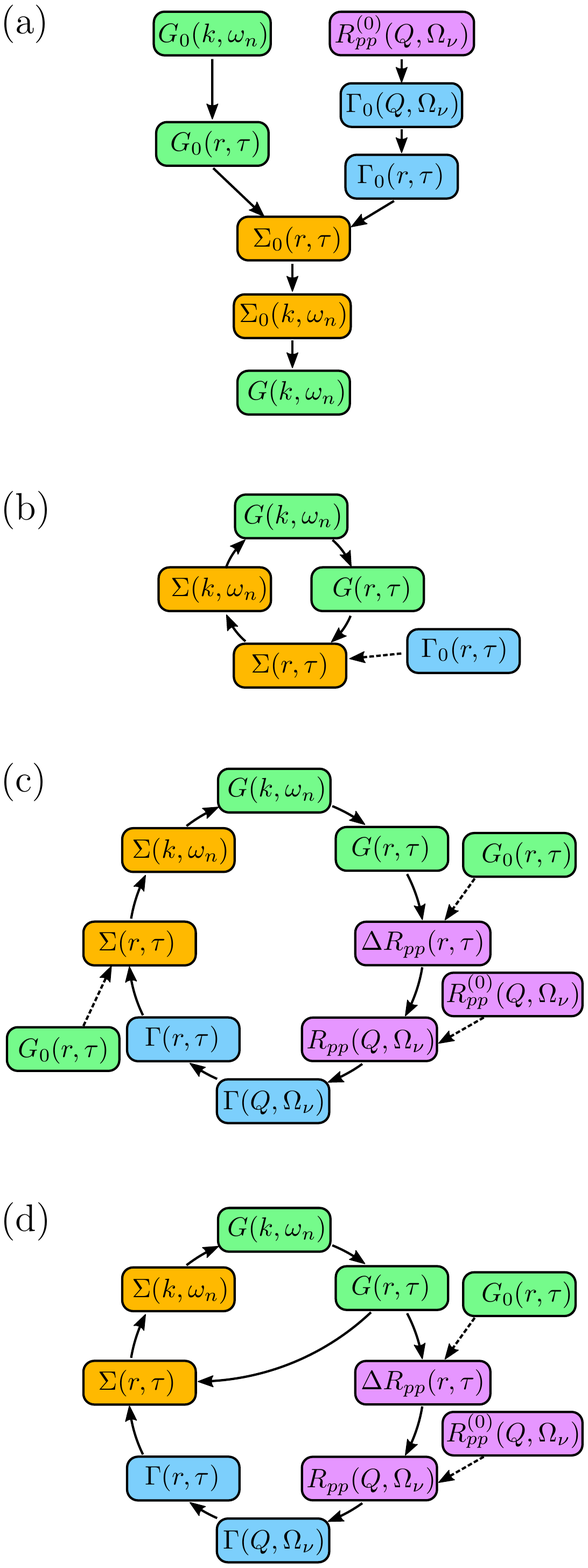}
\caption{(Color online) Flowchart for the routes toward self-consistency of the various $t$-matrix approaches: (a) non-self-consistent $(G_{0} G_{0})G_{0}$; (b) extended $t$-matrix $(G_{0} G_{0})G$; 
                                    (c) partially self-consistent $(GG_{0})G_{0}$ and $(GG)G_{0}$; (d) fully self-consistent $(GG)G$.}
\label{Figure-2}
\end{center}
\end{figure} 

\vspace{0.05cm}
\begin{center}
{\bf B. Routes toward self-consistency}
\end{center}

Except for the non-self-consistent $(G_{0} G_{0})G_{0}$ approach, the equations (\ref{equation-G})-(\ref{equation-Rpp_Q}) must be solved in a self-consistent way. 
To this end, we follow the numerical procedure developed originally in Refs.~\cite{Haussmann-1993} and \cite{Haussmann-1994}. 
The procedure makes use of the Fourier transforms from the $(\mathbf{k},\omega_{n})$ and $(\mathbf{Q},\Omega_{\nu})$ space to the $(\mathbf{r},\tau)$ space, according to the expressions:
\begin{eqnarray}
G(\mathbf{r},\tau) & = & \int \!\! \frac{d\mathbf{k}}{(2 \pi)^3} \, T \sum_{n} e^{i (\mathbf{k} \cdot \mathbf{r} - \omega_n \tau)} G(\mathbf{k},\omega_n) 
\label{equation-FT_G}  \\
\Gamma(\mathbf{r},\tau) & = & \int \!\! \frac{d\mathbf{Q}}{(2 \pi)^3} \, T \sum_{\nu} e^{i (\mathbf{k} \cdot \mathbf{r} - \Omega_\nu \tau)} \Gamma(\mathbf{Q},\Omega_\nu) .
\label{equation-FT_Gamma}
\end{eqnarray}
\noindent
Analogous transformations hold for $\Sigma(\mathbf{k},\omega_n)$ and $R_{\mathrm{pp}}(\mathbf{Q},\Omega_\nu)$. 
Here, $\tau$ is the imaginary time which varies in the interval $(0,1/T)$. 
In the $r=(\mathbf{r},\tau)$ space, the equations (\ref{equation-Sigma_k}) and (\ref{equation-Rpp_Q}) acquire the simple form:
\begin{eqnarray}
\Sigma(r) & = & - \Gamma(r) \, G^{(\mathrm{c})}(-r)
\label{equation-Sigma_r} \\
R_{\mathrm{pp}}(r) & = & G^{(\mathrm{a})}(r)G^{(\mathrm{b})}(r)-\Lambda \delta(r) ,
\label{equation-Rpp_r}
\end{eqnarray}
\noindent
where $\Lambda$ is an appropriate regularization constant that depends on the cutoff in the $\mathbf{k}$-integral of Eq.~(\ref{equation-Rpp_Q}) and becomes infinite together 
with that cutoff \cite{Haussmann-1993}. 
To avoid dealing directly with $\Lambda$, in $r$-space it is convenient to work in terms of the difference
\begin{equation}
\Delta R_{\mathrm{pp}}(r)= R_{\mathrm{pp}}(r)-R^{(0)}_{\mathrm{pp}}(r) = G^{(\mathrm{a})}(r)G^{(\mathrm{b})}(r)-G_{0}(r)^{2} ,
\label{equation-Delta_Rpp}
\end{equation}
where $R^{(0)}_{\mathrm{pp}}(r)$ is the regularized particle-particle bubble built on the non-interacting $G_0$. 
Equations (\ref{equation-G}), (\ref{equation-Gamma_Q}), (\ref{equation-Sigma_r}), and (\ref{equation-Delta_Rpp}), together with the Fourier transforms (\ref{equation-FT_G}) and (\ref{equation-FT_Gamma}), 
form a complete set of equations that need to be solved self-consistently.
The flowchart shown in Fig.~\ref{Figure-2} summarizes schematically the various routes toward self-consistency to be followed when adopting the alternative $t$-matrix approaches of Table~\ref{Table-I}.

Although the non-self-consistent $(G_{0} G_{0})G_{0}$ approach does not require any self-consistent cycling, we have reported it in Fig.~\ref{Figure-2}(a) since its calculation has always to be performed at a preliminary level, to the extent that it is also used as input for the self-consistent calculations.
In this approach, $R^{(0)}_{\mathrm{pp}}(Q)$ is directly calculated according to Eq.~(\ref{equation-Rpp_Q}) with two non-interacting propagators $G_{0}$ in the place of $G^{(\mathrm{a})}$ and $G^{(\mathrm{b})}$.
Figure~\ref{Figure-2}(b) shows the flowchart for the $(G_{0}G_{0})G$ approach (referred to as the ``extended $t$-matrix'' approach in Ref.~\cite{Ohashi-2012}), where self-consistency is present only in the external $G$, while the particle-particle propagator $\Gamma$ coincides with its bare counterpart $\Gamma_{0}$ built on $R^{(0)}(Q)$. 
Figure~\ref{Figure-2}(c) shows the flowchart for both the $(GG_{0})G_{0}$ and $(GG)G_{0}$ approaches, where self-consistency enters only the particle-particle propagator 
$\Gamma$, while the external propagator $G^{(\mathrm{c})}$ is a non-interacting $G_{0}$. 
Finally, Fig.~\ref{Figure-2}(d) shows the flowchart for the fully self-consistent $(GG)G$ approach. 
Note that, in the flowcharts of Figs.~\ref{Figure-2}(c) and (d), the regularized particle-particle bubble $R_{\mathrm{pp}} (\mathbf{Q},\Omega_{\nu})$ is obtained by Fourier transforming the difference 
$\Delta R_{\mathrm{pp}}(\mathbf{r},\tau)$ defined in Eq.~(\ref{equation-Delta_Rpp}) and then by adding to it the regularized particle-particle bubble 
$R^{(0)}_{\mathrm{pp}}(\mathbf{Q},\tau)$ of the non-self-consistent approach.

Most functions appearing in the flowcharts of Fig.~\ref{Figure-2} suffer from a slowly decaying tail in the variables $(\mathbf{k},\omega_{n})$ or $(\mathbf{Q},\Omega_{\nu})$, which implies a corresponding singular behavior when $(r,\tau) \rightarrow 0$. 
For this reason, the Fourier transforms should be performed on a logarithmic scale, following the prescriptions given in Refs.~\cite{Haussmann-1994} and \cite{Haussmann-2007}. 
In addition, one also needs to subtract appropriate semi-analytic expressions from the functions to be numerically Fourier transformed, in order to make their slow decay faster 
(or, alternatively, their singular behavior weaker). 
These semi-analytic expressions have to be known also for the transformed representation, so that they can be added back to the functions after having performed the Fourier transform. 
A detailed account of the semi-analytic expressions used in the numerical calculations is given in Appendix~\ref{sec:appendix-A}.

\vspace{0.05cm}
\begin{center}
{\bf C. Thouless criterion}
\end{center}

As already mentioned, all five $t$-matrix approaches considered in this paper will be examined on equal footing throughout the whole BCS-BEC crossover (also with emphasis on the analytic results that can be obtained separately in the BCS and BEC limits).
The BCS-BEC crossover is spanned in terms of the (dimensionless) coupling parameter $(k_{F} a_{F})^{-1}$, where $k_{F}=(3 \pi^{2} n)^{1/3}$ is the Fermi wave vector associated with the particle density $n$.
In practice, the crossover between the BCS and BEC regimes is essentially exhausted within the range $-1 \lesssim (k_{F} a_{F})^{-1} \lesssim +1$ across the unitary limit at $(k_{F}a_{F})^{-1} = 0$ 
(for a recent comprehensive account of the BCS-BEC crossover, see Ref.~\cite{Physics-Reports-2018}).

In the present paper, we are interested in the normal phase above the the critical temperature $T_{c}$ of the superfluid transition, where the numerical value of $T_{c}$ depends on the specific theoretical 
approximation one is adopting to describe the Fermi gas.
For all five $t$-matrix approaches we are considering, $T_{c}$ is determined by the Thouless criterion \cite{Thouless-1960}, in the form:
\begin{equation}
\big[ \Gamma(\mathbf{Q}=0,\Omega_\nu=0;T,\mu) \big]^{-1} = 0 .
\label{equation-Thouless_criterion}
\end{equation}
\noindent
This condition has to be supplemented by the density equation to determine the chemical potential $\mu$
\begin{equation}
n = - 2 \, G(\mathbf{r}=0, \tau \rightarrow \beta^{-};T,\mu) \,\,\,\, 
\label{equation-density_equation}
\end{equation}
\noindent
(with $\beta = 1/T$ the inverse temperature), where the factor of $2$ accounts for the spin multiplicity.
In practice, one fixes the values of the coupling $(k_{F} a_{F})^{-1}$ and of the temperature $T$ to determine $\mu$ from Eq.~(\ref{equation-density_equation}), and then uses this value of $\mu$ to determine $T_{c}$
from Eq.~(\ref{equation-Thouless_criterion}).
However, for temperatures close to $T_{c}$ this simple iterative procedure may not work properly as far as the $(GG_{0})G_{0}$, $(GG)G_{0}$ and $(GG)G$ approaches are concerned, where difficulties are found 
in the convergence of the iterative procedure toward self-consistency.
To overcome these difficulties, we have found it necessary to introduce two different refinements of the above iterative procedure, the first of which enables us to get close to $T_{c}$ from $T>T_{c}$ while 
the second one allows us to work exactly at $T=T_{c}$. 
These refinements are described in detail in Appendix~\ref{sec:appendix-B}.

\section{Numerical results for thermodynamic quantities} 
\label{sec:numeric-termodynamic}

Several thermodynamic quantities of interest can be obtained directly in terms of the single-particle fermionic propagator $G$ of Eq.~(\ref{equation-G}) and of the particle-particle propagator $\Gamma$ of 
Eq.~(\ref{equation-Gamma_Q}).
Here, we report on the numerical results obtained within the five alternative $t$-matrix approaches that we are considering, for the critical temperature and chemical potential as well as for the Tan's contact.

\vspace{0.05cm}
\begin{center}
{\bf A. Critical temperature}
\end{center}

\begin{figure}[t]
\begin{center}
\includegraphics[width=8.5cm,angle=0]{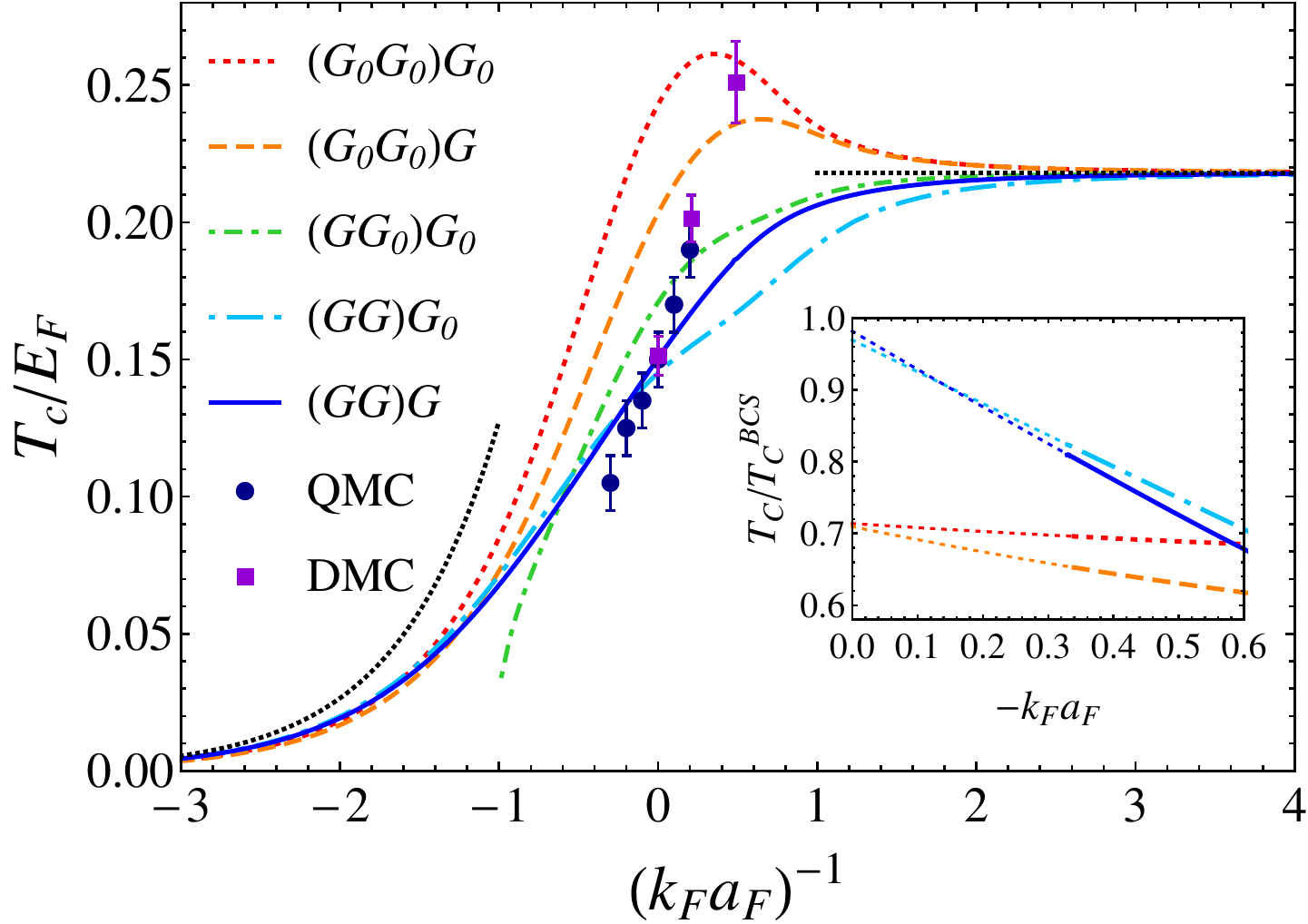}
\caption{(Color online) Critical temperature $T_{c}$ (in units of the Fermi temperature $T_{F}$) as a function of $(k_{F} a_{F})^{-1}$ within alternative $t$-matrix approaches. 
                                   The black dotted lines correspond to the BCS critical temperature (\ref{equation-TcBCS}) when $(k_F a_F)^{-1} < -1$ and to the critical temperature 
                                   (\ref{equation-TcBEC}) for a condensate of non-interacting composite bosons when $(k_F a_F)^{-1} > 1$. 
                                   Quantum Monte Carlo (QMC from Ref.~\cite{Bulgac-2008} - circles) and Diagrammatic Monte Carlo (DMC from Ref.~\cite{Burovski-2008} - squares) data 
                                   are also shown for comparison.
                                    The inset shows the extrapolation of the ratio $T_{c}/T^\text{BCS}_c$ in the weak-coupling limit $(k_{F} a_{F})^{-1} \ll -1$. 
                                    Here, thick lines correspond to numerical data and thin dotted lines to parabolic fits of the data that extrapolate $T_{c}$ in the weak-coupling limit.}
\label{Figure-3}
\end{center}
\end{figure} 

The results for the critical temperature $T_{c}$ obtained from the Thouless criterion (\ref{equation-Thouless_criterion}) are shown in Fig.~\ref{Figure-3} over a wide coupling range for all the five $t$-matrix approaches reported in Table~\ref{Table-I}. 
Several interesting features can be highlighted when comparing the results of the various approaches:

\noindent
(i) All approaches are seen to interpolate rather well between the BCS and BEC critical temperatures (indicated by black dotted lines in Fig.~\ref{Figure-3}), with the notable exception of the $(GG_{0}) G_{0}$ approach which fails to reach the BCS limit since in this case the critical temperature collapses abruptly to zero at coupling $(k_{F} a_{F})^{-1} \simeq -1$. 
We attribute this failure to the asymmetric treatment of the single-particle propagators $G^{(\mathrm{a})}$ and $G^{(\mathrm{b})}$ in the particle-particle bubble (\ref{equation-Rpp_Q}) that enters the particle-particle propagator $\Gamma$. 
The asymmetry generates an artificial imbalance between spin up and down species, which acts to suppress the critical temperature of the superfluid transition (cf. also Appendix~\ref{sec:appendix-C}).
This feature has apparently passed unnoticed in the literature.
It is for this reason that, in what follows, the results of the $(GG_{0}) G_{0}$ approach will not be reported for the BCS limit.

\noindent
(ii) The behavior of the critical temperature in the BCS (weak-coupling) limit $(k_{F} a_{F})^{-1} \ll -1$ is shown as a function of $k_{F} a_{F}$ in the inset of Fig.~\ref{Figure-3}), 
for all $t$-matrix approaches with the exception of the $(GG_{0}) G_{0}$ approach for the reasons discussed in point (i) above.
In all cases, the numerical results have been extended toward the \emph{extreme} BCS limit $k_{F} a_{F} \rightarrow 0^{-}$ through a parabolic extrapolation (dots).
Within a numerical error of the order of $1\%$ \cite{footnote-1}, in this limit we obtain that the $(GG)G$ and $(GG)G_{0}$ approaches recover the value of the BCS temperature:
\begin{equation}
T_{c}^\text{BCS} = \frac{ 8 e^{\gamma} E_{F}}{\pi e^{2}}\exp \Big(\frac{\pi}{2 k_{F} a_{F}}\Big)
\label{equation-TcBCS}
\end{equation}
where $\gamma$ is the Euler's constant. 
On the other hand, the $(G_{0}G_{0})G$ and $(G_{0}G_{0})G_{0}$ approaches reproduce the BCS critical temperature only within logarithmic accuracy, in the sense that they recover 
the result (\ref{equation-TcBCS}) with the pre-factor divided by $e^{1/3}$ (see Ref.~\cite{Pisani-2018-I} for a discussion of the origin of this spurious factor in those approaches that do not dress 
the bare particle-particle propagator $\Gamma_{0}$).

\noindent
(iii) In the crossover region about unitarity, it turns out that the inclusion of (even a partial degree of) self-consistency in the particle-particle propagator $\Gamma$ of 
Eq.~(\ref{equation-Gamma_Q}) acts to suppress the maximum of $T_{c}$, which otherwise occurs for the $(G_{0}G_{0})G$ and $(G_{0}G_{0})G_{0}$ approaches.

\noindent
(iv) In the BEC (strong-coupling) limit $(k_F a_F)^{-1} \gg 1$, all approaches reproduce the value of the critical temperature for a condensate of non-interacting composite bosons made up of fermion pairs (with mass $m_{B}=2m$ and density $n_{B}=n/2$):
\begin{equation}
T_{c}^\text{BEC} = \frac{2 \pi}{\zeta(3/2)^{2/3}} \frac{\big(n_{B}\big)^{2/3}}{m_{B}} \simeq 0.218 \, E_{F} .
\label{equation-TcBEC}
\end{equation}
However, alternative $t$-matrix approaches differ in the way the value (\ref{equation-TcBEC}) is reached when $(k_F a_F)^{-1} \gg 1$.
Specifically, the sub-leading behavior of $T_{c}$ when approaching $T_{c}^\text{BEC}$ can be characterized by the expression
\begin{equation}
\frac{T_{c} -T^\text{BEC}_{c}}{T^\text{BEC}_{c}} = \frac{\alpha}{3 \pi} (k_{F} a_{F})^{3} ,
\label{subleading-BEC}
\end{equation}
where the values of the coefficient $\alpha$ for the various $t$-matrix approaches are listed in Table~\ref{Table-II}.

\begin{table}[h]
\begin{tabular}{ ccc } 
\hline
\hline
$(G^{(a)} G^{(b)})G^{(c)}$ & $\alpha$(th)  & $\alpha$(extr) \\ 
\hline
 $(G_{0}G_{0})G_{0}$ & 1 & 1.02   \\ 
 $(G_{0}G_{0})G$ & 1.07 & 1.12    \\ 
 $(GG_{0})G_{0}$ & -0.5 & -0.48    \\ 
 $(GG)G_{0}$ & na & -2.00            \\ 
 $(GG)G$ & -1 & -1.02                   \\ 
\hline
\hline
\end{tabular}
\caption{Values of the coefficient $\alpha$ of Eq.~(\ref{subleading-BEC}) obtained within the various $t$-matrix approaches, both from analytic calculations (th) and extrapolation of numerical results (extr). One analytic value is not available (na).}
\label{Table-II}
\end{table}

The theoretical values reported in Table~\ref{Table-II} for the $(GG)G$ and $(GG_{0})G_{0}$ approaches are taken from Refs.~\cite{Haussmann-1993} and \cite{Chen-2005}, respectively, while we have calculated independently those for the $(G_{0} G_{0})G_{0}$ and $(G_{0} G_{0})G$ approaches (the corresponding analytic calculations are not reported here owing to their complexity).
Note that, for those approaches that include even a partial degree of self-consistency in the particle-particle propagator $\Gamma$, $\alpha < 0$ such that the value 
(\ref{equation-TcBEC}) is approached from below; the opposite occurs for the remaining approaches. 

We have also compared the numerical values we have obtained for $T_{c}$ within the various $t$-matrix approaches with other published data for the same quantity.
For the $(G_{0} G_{0})G_{0}$, $(G_{0} G_{0})G$, and $(GG)G$ approaches we found good agreement with the data published in Refs.~\cite{PPSC-2002}, \cite{Ohashi-2018}, and \cite{Haussmann-1994}, respectively.
For the $(GG_{0})G_{0}$ and $(GG)G_{0}$ approaches, on the other hand, direct comparison with previous data throughout the BCS-BEC crossover is not possible, since within these approaches
the curves for $T_{c}$ have only been calculated with additional approximations affecting the form of the particle-particle propagator and of the self-energy \cite{Levin-2010} \cite{Micnas-2014}. 
In these cases, our data compares well with those of Refs.~\cite{Levin-2010} and \cite{Micnas-2014} only in the strong-coupling regime $(k_F a_F)^{-1} \gtrsim 1$, where the additional approximations introduced in those references remain valid at $T \simeq T_{c}$ (see also Appendix~\ref{sec:appendix-C}).

Finally, a comparison with available quantum Monte Carlo (both QMC and DMC) data has been reported in Fig.~\ref{Figure-3}. Note how these data show a steeper coupling dependence with respect to the $t$-matrix calculations. As already mentioned in the Introduction, this steeper dependence can be accounted for by a further inclusion of the GMB vertex corrections \cite{Pisani-2018-I}.

\vspace{0.05cm}
\begin{center}
{\bf B. Chemical potential}
\end{center}

\begin{figure}[t]
\begin{center}
\includegraphics[width=8.5cm,angle=0]{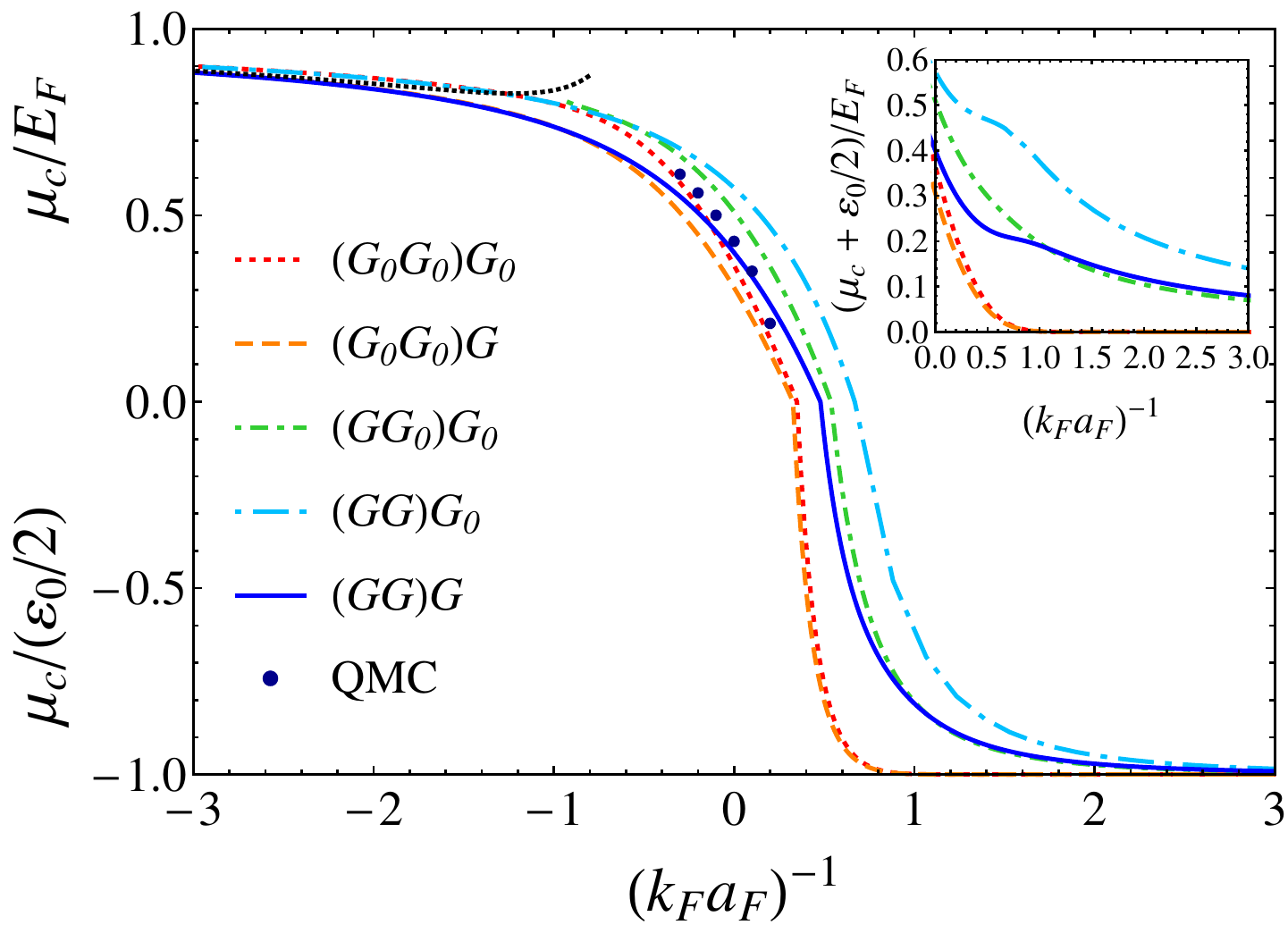}
\caption{(Color online) Chemical potential $\mu_{c}$ at $T_{c}$ (in units of the Fermi energy $E_{F}$ when $\mu_{c}>0$ and of half the binding energy 
                                     $\varepsilon_{0}=(m a_{F}^{2})^{-1}$ of composite bosons when $\mu_{c}<0$) as a function of $(k_{F} a_{F})^{-1}$ within alternative $t$-matrix approaches.
                                     The black dotted line in weak coupling corresponds to the Galitskii's result (\ref{equation-muc_Galitskii}).
                                     Comparison with QMC data from Ref.~\cite{Bulgac-2008} (circles) is also shown.
                                     In the inset, $\varepsilon_{0}/2$ has been added to $\mu_{c}$ for $(k_{F} a_{F})^{-1}>0$ to amplify the values of the chemical potential of composite bosons 
                                     due to their mutual interaction.}
\label{Figure-4}
\end{center}
\end{figure} 

The corresponding results for the chemical potential $\mu_{c}$ calculated at the critical temperature $T_{c}$ are reported in Fig.~\ref{Figure-4} as a function of coupling 
$(k_{F} a_{F})^{-1}$, for each of the five $t$-matrix approaches here considered.
The main features that can be identified from this plot are:

\noindent
(i) In the weak-coupling (BCS) limit $(k_{F} a_{F})^{-1} \ll -1$, the $(GG)G$ and $(G_0G_0)G$ approaches recover the Galitskii's expression \cite{Galitskii-1958}
\begin{equation}
\frac{\mu^\text{Gal}}{E_{F}} = 1+\frac{4}{3\pi}(k_{F} a_{F})+\frac{4}{15 \pi^2}\big[11-2 \ln2 \big] (k_{F} a_{F})^2 
\label{equation-muc_Galitskii}
\end{equation}
up to second order in $(k_{F} a_{F})$, while the $(G_{0}G_{0})G_{0}$ and $(G_{0}G_{0})G$ approaches recover this expression up to first order only. 

\noindent
(ii) In the strong-coupling (BEC) limit $(k_F a_F)^{-1} \gg 1$, all approaches recover the value $\mu_{c} = -\varepsilon_{0}/{2}$ at leading order in $k_{F} a_{F}$, 
where $\varepsilon_{0}=(m a_{F}^{2})^{-1}$ is the binding energy of the composite bosons. 
At sub-leading order, on the other hand, the behavior of $\mu_{c}$ depends on the approach.
In particular, for the $(G_{0}G_{0})G_{0}$ and $(G_{0}G_{0})G$ approaches the sub-leading term vanishes exponentially when $(k_{F} a_{F})^{-1} \gg 1$, while for the $(GG_{0})G_{0}$,
$(GG)G_{0}$, and $(GG)G$ approaches the sub-leading term vanishes linearly in $(k_{F} a_{F})$ (a behavior which has been evidenced in the inset of Fig.~\ref{Figure-4}). 
This difference is due to the fact that self-consistency in the particle-particle propagator introduces a residual repulsive interaction between composite bosons, as discussed in subsection~\ref{sec:numeric-termodynamic}-C.

\begin{figure}[h]
\begin{center}
\includegraphics[width=8.0cm,angle=0]{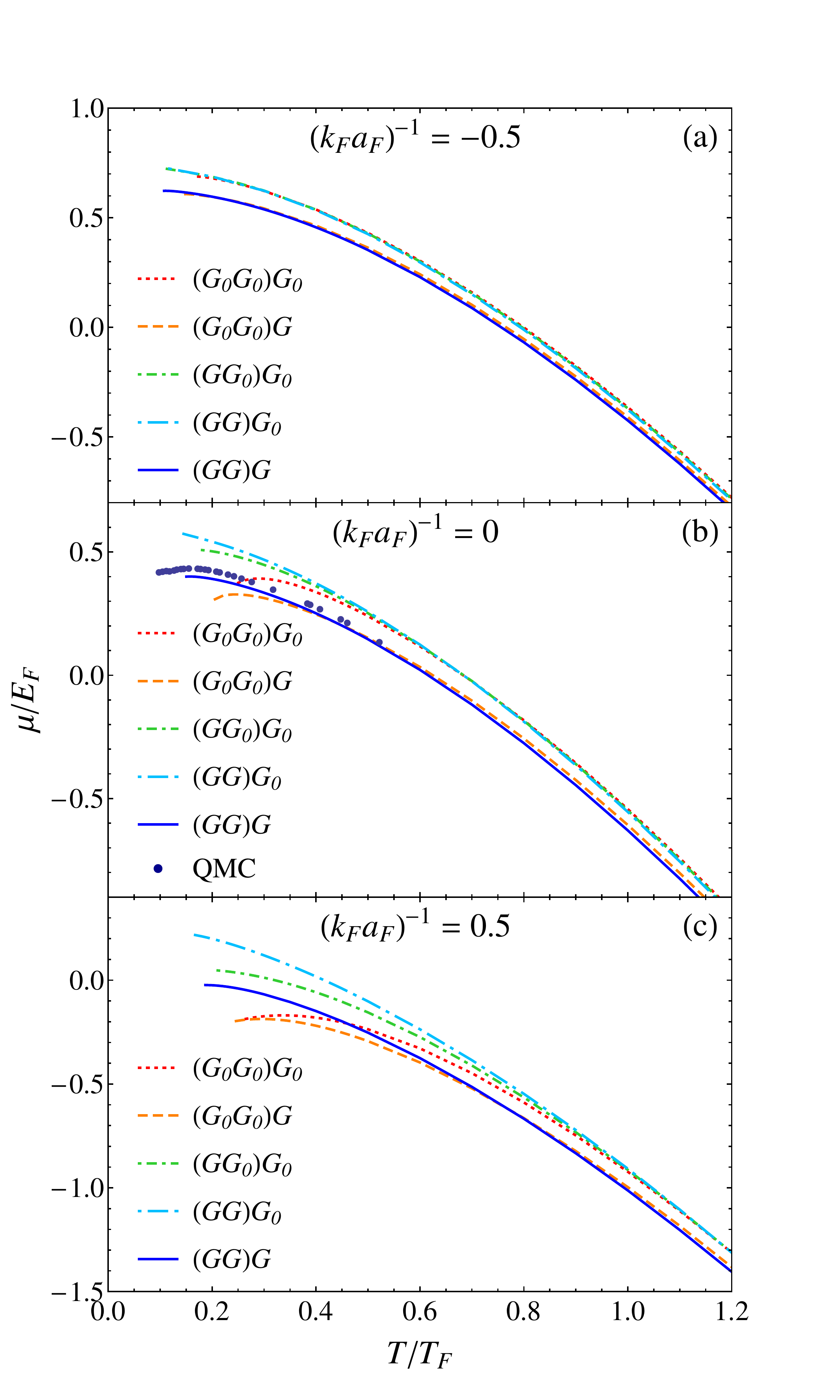}
\caption{(Color online) Chemical potential $\mu$ (in units of $E_{F}$) as a function of the temperature $T$ (in units of $T_{F}$) for various couplings $(k_{F} a_{F})^{-1} = (-0.5, 0, 0.5)$           
                                     within alternative $t$-matrix approaches.
                                     Comparison with QMC data from Ref.~\cite{Drut-2012} (circles) is also shown in panel (b).}
\label{Figure-5}
\end{center}
\end{figure} 

By the present methods, the chemical potential $\mu$ can be calculated not only at $T_{c}$ but also \emph{above} $T_{c}$.
Figure~\ref{Figure-5} shows $\mu$ as a function of temperature for three characteristic couplings [$(k_{F} a_{F})^{-1} = (-0.5, 0, 0.5)$] spanning the crossover regime, for all five $t$-matrix approaches we are considering.
On the low-$T$ side, the curves terminate at the respective critical temperature $T_{c}$ given in Fig.~\ref{Figure-3}.
Note that, for the $(G_{0}G_{0})G_{0}$ and $(G_{0}G_{0})G$ approaches where no degree of self-consistency is introduced in the particle-particle propagator, the curves of $\mu(T)$ show 
a maximum above $T_{c}$.
We have also verified that the curve of Fig.~\ref{Figure-5}(b), which corresponds to the $(GG_{0})G_{0}$ approach at unitarity, compares well with the numerical results reported in Ref.~\cite{Drummond-2008}.
To the best of our knowledge, this is, in fact, the only other reference where a calculation based on the \emph{complete} $(GG_{0})G_{0}$ approach was performed, without recourse to additional simplifying approximations (albeit in Ref.~\cite{Drummond-2008} the calculation was limited to the coupling $(k_{F} a_{F})^{-1} = 0$ only).

In both Figs.~\ref{Figure-4} and \ref{Figure-5}, comparison has also been added with available QMC data, which show overall good agreement with the results of the $t$-matrix approaches.

\vspace{0.05cm}
\begin{center}
{\bf C. Scattering length of composite bosons}
\end{center}

\begin{figure}[t]
\begin{center}
\includegraphics[width=7.5cm,angle=0]{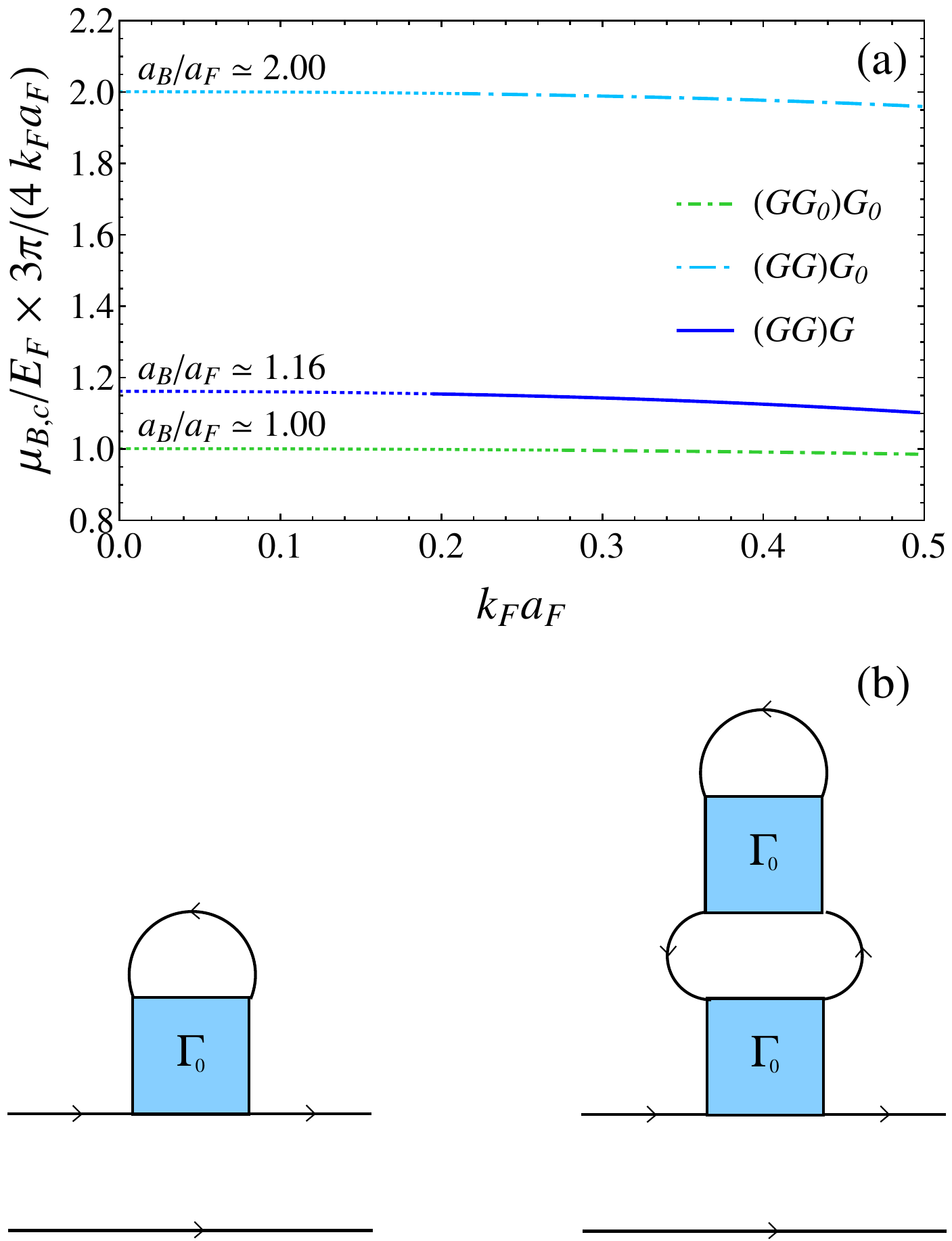}
\caption{(Color online) (a) Values of $a_{B}/a_{F}$ extrapolated from the expression (\ref{equation-aB_extrapolation}) within alternative $t$-matrix approaches. 
                                     Thick lines correspond to numerical data, while thin dotted lines correspond to parabolic fits to numerical data
                                      which extrapolate to the value of $a_{B}/a_{F}$ in the strong-coupling limit $k_{F} a_{F} \rightarrow 0^{+}$. 
                                      (b) Diagrams contributing to the bosonic scattering length $a_{B}$ within the $(GG)G$ approach.
                                      Like in Fig.~\ref{Figure-1}, thin lines correspond to the bare fermionic propagator $G_{0}$.}
\label{Figure-6}
\end{center}
\end{figure}

The residual repulsive interaction between composite bosons is characterized by a finite value of the scattering length $a_{B}$.
This can, in turn, be determined by comparing the chemical potential for the composite bosons $\mu_{B} = 2 \mu_{c}+\varepsilon_{0}$ in the strong-coupling limit $(k_{F} a_{F})^{-1} \gg 1$ with the chemical potential $\mu^{0}_{B}=8\pi a_{B} n_{B}/m_{B}$ of a dilute Bose gas at $T_{c}$.
This comparison yields:
\begin{equation}
\frac{a_{B}}{a_{F}} = \lim_{k_{F} a_{F} \to 0^+} \ \frac{3 \pi}{4 (k_{F} a_{F})}  \frac{\mu_{B}}{E_F} \, .
\label{equation-aB_extrapolation}
\end{equation}
The values of $a_{B}$ extrapolated in this way for the $(GG)G$, $(GG)G_{0}$ and $(GG_{0})G_{0}$ approaches are shown in Fig.~\ref{Figure-6}(a).
Note, in particular, that the $(GG)G$ fully self-consistent approach yields $a_{B}/a_{F} \simeq 1.16$, which contrasts with the value $a_{B}/a_{F}=2$ obtained analytically in Ref.~\cite{Haussmann-1993}.

The above limiting numerical values for the bosonic scattering length $a_{B}$, obtained within the $t$-matrix approaches that include the dressed single-particle propagator $G$
in the particle-particle propagator $\Gamma$, can be compared with the corresponding analytic results obtained by expressing the self-consistently dressed $\Gamma$ as an infinite series of diagrams in terms 
of the bare $G_{0}$ and $\Gamma_{0}$, and then by retaining only the leading-order corrections to the bare particle-particle propagator $\Gamma_{0}$ in the BEC ($\mu/T \to -\infty$) limit.

For the $(GG)G_{0}$ approach, this corresponds to considering the left diagram of Fig.~\ref{Figure-6}(b) (together with the corresponding diagram where is the lower fermionic line to get dressed). 
Comparison of the analytic evaluation of this diagram in the BEC limit as done in Refs.~\cite{Haussmann-1993} and \cite{Pieri-2000} with the leading self-energy correction $\Sigma_{B} = 8 \pi a_B n_B / m_B$ for a dilute Bose gas  yields the value $a_{B}/a_{F} = 2$ (cf. the upper curve of Fig.~\ref{Figure-6}(a)).

For the $(GG_{0})G_{0}$ approach, on the other hand, the absence of the self-energy correction in the lower fermionic line then eliminates the multiplicity of $2$ for the left diagram of Fig.~\ref{Figure-6}(b), yielding $a_{B}/a_{F}=1$ (cf.~the lower curve of Fig.~\ref{Figure-6}(a))
Note, however, that this reduction of $a_{B}$ by half in comparison with the more symmetric $(GG)G_{0}$ approach is somewhat artificial, since it corresponds to an incomplete symmetrization of the bosonic interaction vertex. 
   
Finally, for the $(GG)G$ approach also the right diagram of Fig.~\ref{Figure-6}(b) contributes at the leading order, again with a multiplicity of $2$ due to the corresponding dressing 
of the lower fermionic line.  
This diagram (which had apparently escaped to the analysis of Ref.~\cite{Haussmann-1993}) yields a correction $-0.842 a_{F}$ to the bosonic scattering length $a_{B}$ in the BEC limit, thereby resulting altogether in the value $a_{B}/a_{F}=2-0.842=1.158 $ in excellent agreement (within $0.2 \%$) with that obtained by our numerical extrapolation (cf. the middle curve of Fig.~\ref{Figure-6}(a)).  
This analytic estimate for the right diagram of Fig.~\ref{Figure-6}(b) can be readily obtained by noting that, in the BEC limit, its leading-order behavior coincides with that of the diagram introduced in Ref.~\cite{Pisani-2018-I} to include the GMB correction for $T_{c}$ throughout the BCS-BEC crossover, whose contribution to $a_{B}$ in the BEC limit was there estimated to be $-0.842 a_{F}$.  
It should also be remarked that this identification between (the numerical values of) the two diagrams holds in the BEC limit only.
For this reason, the self-consistent $t$-matrix approximation fails to recover the GMB reduction factor for $T_{c}$ in the BCS limit \cite{Pisani-2018-I}.

\vspace{0.05cm}
\begin{center}
{\bf D. Tan's contact}
\end{center}

\begin{figure}[t]
\begin{center}
\includegraphics[width=8.5cm,angle=0]{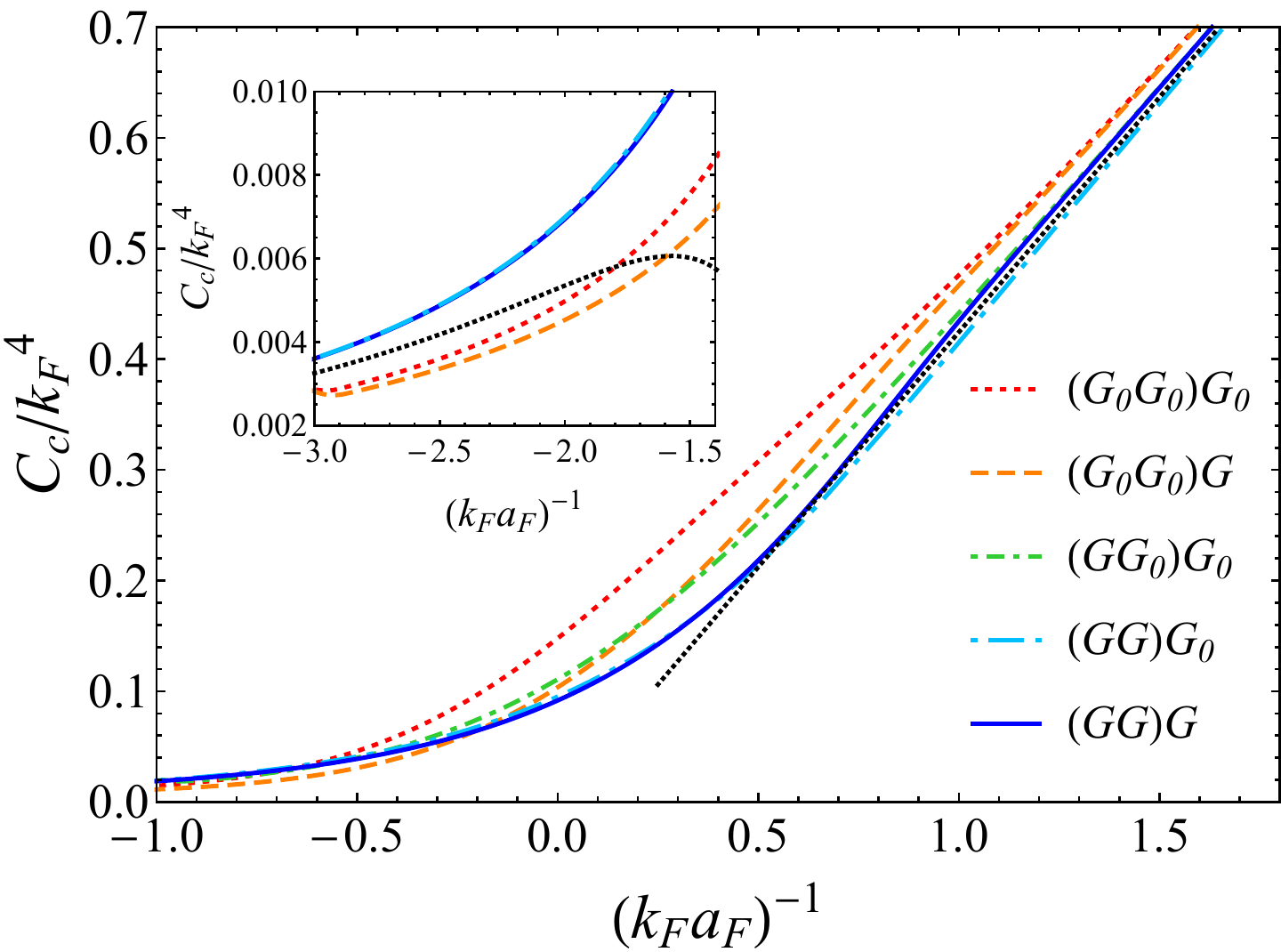}
\caption{(Color online) The contact $C_{c}$ at $T_{c}$ is shown as a function of $(k_{F} a_{F})^{-1}$ within alternative $t$-matrix approaches. 
                                    The black dotted line stands for the leading term of the contact $C_{c}/k_{F}^{4}=4/(3\pi k_{F} a_{F})$ in the strong-coupling limit (cf. Ref.~\cite{Physics-Reports-2018}). 
                                    The inset reports the contact on the weak-coupling side, where the black dotted line now corresponds to the expression (\ref{equation-C_Galitskii}) 
                                    within the Galiskii's approximation up to next-to-leading order in $(k_{F} a_{F})$.}
\label{Figure-7}
\end{center}
\end{figure}

An important physical quantity that characterizes a Fermi gas with short-range interaction is the Tan's contact \cite{Tan-2008-I,Tan-2008-II,Tan-2008-III}, which connects two-particle correlations 
at short distances with thermodynamics.
Here, we calculate the contact $C_{c}$ at $T_{c}$ within the various $t$-matrix approaches as the trace of the particle-particle propagator
\begin{equation}
C_{c} = \int \!\! \frac{d\mathbf{Q}}{(2\pi)^3} T \sum_\nu \Gamma(\mathbf{Q},\Omega_\nu) e^{i \Omega_{\nu} 0^{+}} = \Gamma(\mathbf{r}=0, \tau \rightarrow \beta^{-}) ,
\label{equation-contact}
\end{equation}
according to an expression introduced in Ref.~\cite{PPS-NP-2009}.
The results are shown in Fig.~\ref{Figure-7} throughout the BCS-BEC crossover.
For internal consistency, we have also verified numerically that the values of the contact $C_{c}$ obtained by Eq.~(\ref{equation-contact}) coincide with the coefficient of the $\mathbf{k}^{-4}$ tail of the wave-vector distribution (per spin component $\sigma$)
\begin{equation}
n_{\sigma}(\mathbf{k}) = - \, G(\mathbf{k}, \tau \rightarrow \beta^{-}) . 
\label{equation-wave-vector-distribution}
\end{equation}

We have further verified that, at $T_{c} \, (\ll T_{F})$ in weak coupling, the expansion for the contact in powers of $(k_{F} a_{F})$
\begin{equation}
\frac{C_{c}}{k_F^4}=\frac{4(k_{F} a_{F})^2}{9\pi^2}\bigg(1+\frac{12}{35 \pi}(11-2\ln2)(k_{F} a_{F})\bigg) ,
\label{equation-C_Galitskii}
\end{equation}
that results by taking the derivative with respect to $a_{F}^{-1}$ of the expression for the total energy at $T=0$ obtained by Galitskii in powers of $(k_{F} a_{F})$ \cite{Galitskii-1958}, 
is recovered by the $(G_{0}G_{0})G_{0}$ and $(G_{0}G_{0})G$ approaches up to the leading order and by the $(GG)G$ and $(GG)G_{0}$ approaches up to the next-to-leading order in $(k_{F} a_{F})$.

\vspace{0.05cm}
\begin{center}
{\bf E. Summary of the main thermodynamic results}
\end{center}

\begin{table}[t]
\begin{tabular}{ cccc } 
\hline
 \hline
$(k_{F} a_{F})^{-1}$ & $T_{c}/T_{F}$ & $\mu_{c}/E_{F}$ & $C_{c}/k_{F}^{4}$ \\
\hline
\hline
  & 0.08495 & 0.7997 & 0.01462 \\ 
 & 0.07294 & 0.7381 & 0.01139\\
 -1.0& na & na & na \\
  & 0.07132 & 0.8006 & 0.01930 \\ 
 & 0.06776& 0.7375 & 0.01874 \\
 \hline
  & 0.1649 & 0.6846 & 0.04595 \\ 
 & 0.1358 & 0.6018 & 0.03076 \\
 -0.5& 0.1132 & 0.7230 & 0.03999 \\
  & 0.1112 & 0.7269 & 0.04099 \\ 
 & 0.1077 & 0.6226 & 0.03892\\
 \hline
 & 0.2429 & 0.3655 & 0.1479\\ 
 & 0.2034 & 0.3059 & 0.1036 \\
 0.0 & 0.1709 & 0.5096 & 0.1108\\
 & 0.1451 & 0.5734 & 0.09513 \\
 & 0.1505  & 0.4000 & 0.09170\\
 \hline
   & 0.2588 & -0.1890 & 0.3075 \\ 
 & 0.2361 & -0.2005 & 0.2638 \\
 0.5& 0.1975 & 0.04935 & 0.2522 \\
  & 0.1674 &  0.2176 & 0.2148 \\ 
 & 0.1870  &  -0.02339 &  0.2182 \\
 \hline
   & 0.2351 & -0.9986& 0.4759 \\ 
 & 0.2320& -0.9987 & 0.4666 \\
 1.0& 0.2097 & -0.8042 & 0.4418 \\
  & 0.1918 & -0.6250 & 0.4150 \\ 
 & 0.2062 & -0.8092 & 0.4342 \\
\hline
\hline
\end{tabular}
\caption{Numerical values of the critical temperature $T_{c}$, chemical potential $\mu_{c}$, and contact $C_{c}$ at $T_{c}$ are reported for five characteristic 
             couplings in the crossover region about unitarity and for all $t$-matrix approaches. 
             A few numerical values are not available (na) for the reasons discussed in the text.
             For each coupling, reference to different $t$-matrix approaches follows the conventions (from top to bottom) of Table~\ref{Table-I}.}
\label{Table-III}
\end{table}

The numerical values of the critical temperature $T_{c}$, and of the chemical potential $\mu_{c}$ and the contact $C_{c}$ at $T_{c}$, can be read off directly from 
Figs.~\ref{Figure-3}, \ref{Figure-4}, and \ref{Figure-7}, respectively, for each of the five $t$-matrix approaches that we have considered in this paper.
It might be useful, however, to summarize the values of the above physical quantities for a few characteristic couplings in the crossover region of most interest.
Accordingly, we report in Table~\ref{Table-III} a list of these values for five couplings in the interval $-1 \le (k_{F} a_{F})^{-1} \le +1$.
In this way, differences in the results of the various $t$-matrix approaches can be most readily appreciated.

\section{Numerical results for dynamical quantities} 
\label{sec:numeric-dynamic}

The results obtained in Section~\ref{sec:numeric-termodynamic}, about the single-particle fermionic propagator $G$ of Eq.~(\ref{equation-G}), can now be utilized to obtain a number of spectral features for the attractive Fermi gas in the normal phase throughout the BCS-BEC crossover.
To this end, a suitable method is required to perform the analytic continuation from Matsubara to real frequencies, as discussed next.

\vspace{0.05cm}
\begin{center}
{\bf A. Method for analytic continuation}
\end{center}

To assess the dynamical properties of a Fermi gas, the analytic continuation $\bar{G}(\mathbf{k},z)$ of the single-particle fermionic propagator $G(\mathbf{k},\omega_{n})$ is required over the complex $z$-plane. 
The function $\bar{G}(\mathbf{k},z)$ takes the values $G(\mathbf{k},\omega_{n})$ along the imaginary axis at the Matsubara frequencies $z=i \omega_{n}$, and is analytic everywhere except on the  real frequency axis where its imaginary part is discontinuous due to the time-reversal symmetry condition $\bar{G}(\mathbf{k},z^*)=\bar{G}(\mathbf{k},z)^*$. 
The \emph{single-particle spectral function} $A(\mathbf{k},\omega)$ is then obtained in terms of the discontinuity of $\bar{G}(\mathbf{k},z)$ across the real frequency axis:
\begin{eqnarray}
A(\mathbf{k},\omega) & = &- \frac{1}{2\pi i} \big[ \bar{G}(\mathbf{k},\omega+i 0^{+}) -\bar{G}(\mathbf{k},\omega-i 0^{+}) \big] 
\nonumber \\
& = & - \frac{1}{\pi} \text{Im} \big[ \bar{G}(\mathbf{k},\omega+ i 0^{+}) \big] .
\label{equation-Akw}
\end{eqnarray}
The positive definite function $A(\mathbf{k},\omega)$ is normalized according to:
\begin{equation}
\int_{-\infty}^{+\infty} \!\!\! d\omega \, A(\mathbf{k},\omega) = 1 .
\label{equation-Akw_norm}
\end{equation}

To evaluate the single-particle spectral function (\ref{equation-Akw}), a procedure is required to obtain the function $\bar{G}(\mathbf{k},z)$ just above the real frequency axis from the known values $\bar{G}(\mathbf{k},z=i \omega_{n})= G(\mathbf{k},\omega_n)$ on the imaginary axis. 
To this end, we make use of the method of Pad\'e approximants \cite{Serene-1977,Beach-1999} which consists in approximating $\bar{G}(\mathbf{k},z)$ for given $\mathbf{k}$ by a ratio of polynomials, in the form:
\begin{equation}
\bar{G}(\mathbf{k},z) =\frac{ p_1+p_2 \, z + \cdots + p_r z^{r-1}}{q_1+q_2 \, z + \cdots + q_r z^{r-1}+ z^{r}} \, .
\label{Pade}
\end{equation}
Here, the $2 r$ (real) coefficients $\{p_i,q_i; i=1,\cdots,r\}$ are determined from the values of $\bar{G}(\mathbf{k},z)$ at $2 r$ points on the imaginary axis.

This procedure, however, turns out to be quite sensitive to the presence of numerical uncertainties in the values of the input function $G(\mathbf{k},\omega_{n})$, in such a way that the resulting shape of the function $A(\mathbf{k},\omega)$ may turn out to be rather distorted or even to acquire negative values. 
To mitigate the occurrence of this sort of problems, we have followed a prescription proposed in Ref.~\cite{Schott-2016} and averaged over several (typically, 12) runs of analytic continuations with different sets of $2 r$ (typically, $50$) points selected on the imaginary axis, consistently discarding those runs that yield $A(\mathbf{k},\omega)<0$ for some $\omega$ intervals. 
In addition, even though the analytical continuation is performed over the half-plane with $\text{Im}(z)>0$, we have sometimes found it useful to include in the sets of $2 r$ points the first few (from one up to three) frequencies on the negative imaginary axis, as also proposed in Ref.~\cite{Schott-2016}. 

\begin{figure}[t]
\begin{center}
\includegraphics[width=7.1cm,angle=0]{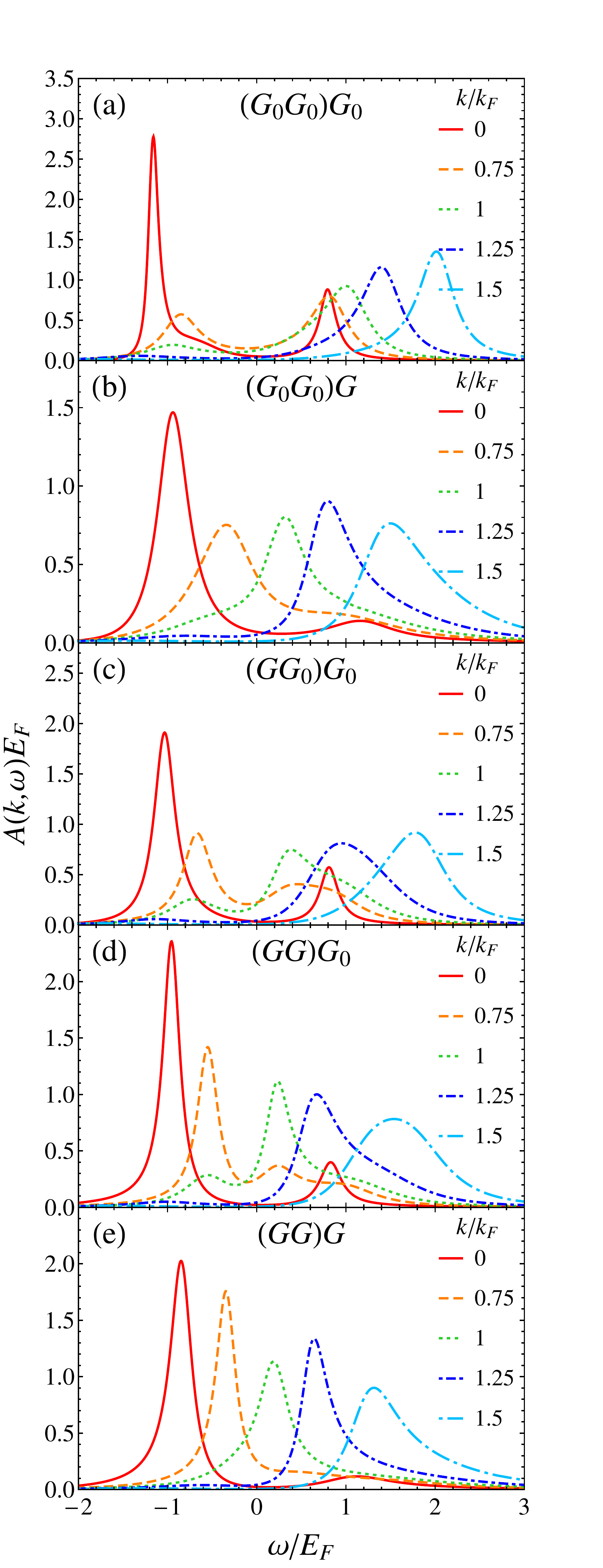}
\caption{(Color online) Spectral function $A(\bm{k},\omega)$ (in units of $E_{F}^{-1}$) at $(k_{F} a_{F})^{-1}=0$ and $T=T_{c}$ for different values of $k$ (in units of $k_{F}$) within alternative $t$-matrix approaches.}
\label{Figure-8}
\end{center}
\end{figure} 

\vspace{0.05cm}
\begin{center}
{\bf B. Single-particle spectral function}
\end{center}

The single-particle spectral function $A(k,\omega)$, obtained at unitarity and $T_{c}$ for all $t$-matrix approaches, is shown in Fig.~\ref{Figure-8} for several values of 
$k = |\mathbf{k}|$. 
A notable difference results by comparing the various panels of Fig.~\ref{Figure-8}, between the approaches that dress the fermionic propagator $G^{(\mathrm{c})}$ in Eq.~(\ref{equation-Sigma_k}) (cf. panels (b) and (e)) and those that do not (cf. panels (a), (c), and (d)). 
As a matter of fact, in the $(GG)G$ and $(G_{0} G_{0})G$ approaches there is essentially no evidence of a double-peak structure (except at $k \simeq 0$), such that the shape of
$A(\mathbf{k},\omega)$ is mostly represented by a single peak that shifts from negative to positive frequencies upon increasing $k$. 
Yet, this behavior appears not to be consistent with what one would expect for a Fermi liquid, as sometimes claimed instead in the literature \cite{Salomon-2011}.
This is because the single peak broadens up just at $k \simeq k_{F}$ with a width of the order of $E_{F}$, both features being not consistent with the behavior of a Fermi liquid \cite{AGD-1975,Nozieres-1964}. 
On the other hand, the $(GG)G_{0}$, $(GG_{0})G_{0}$ and $(G_{0} G_{0})G_{0}$ approaches (whereby $G^{(c)}$ in Eq.~(\ref{equation-Sigma_k}) remains $G_{0}$) present a persistent double-peak structure through $k \simeq k_{F}$, with an exchange of weight occurring for increasing $k$ between the peaks at negative and positive frequencies. 

\begin{figure}[t]
\begin{center}
\includegraphics[width=7.5cm,angle=0]{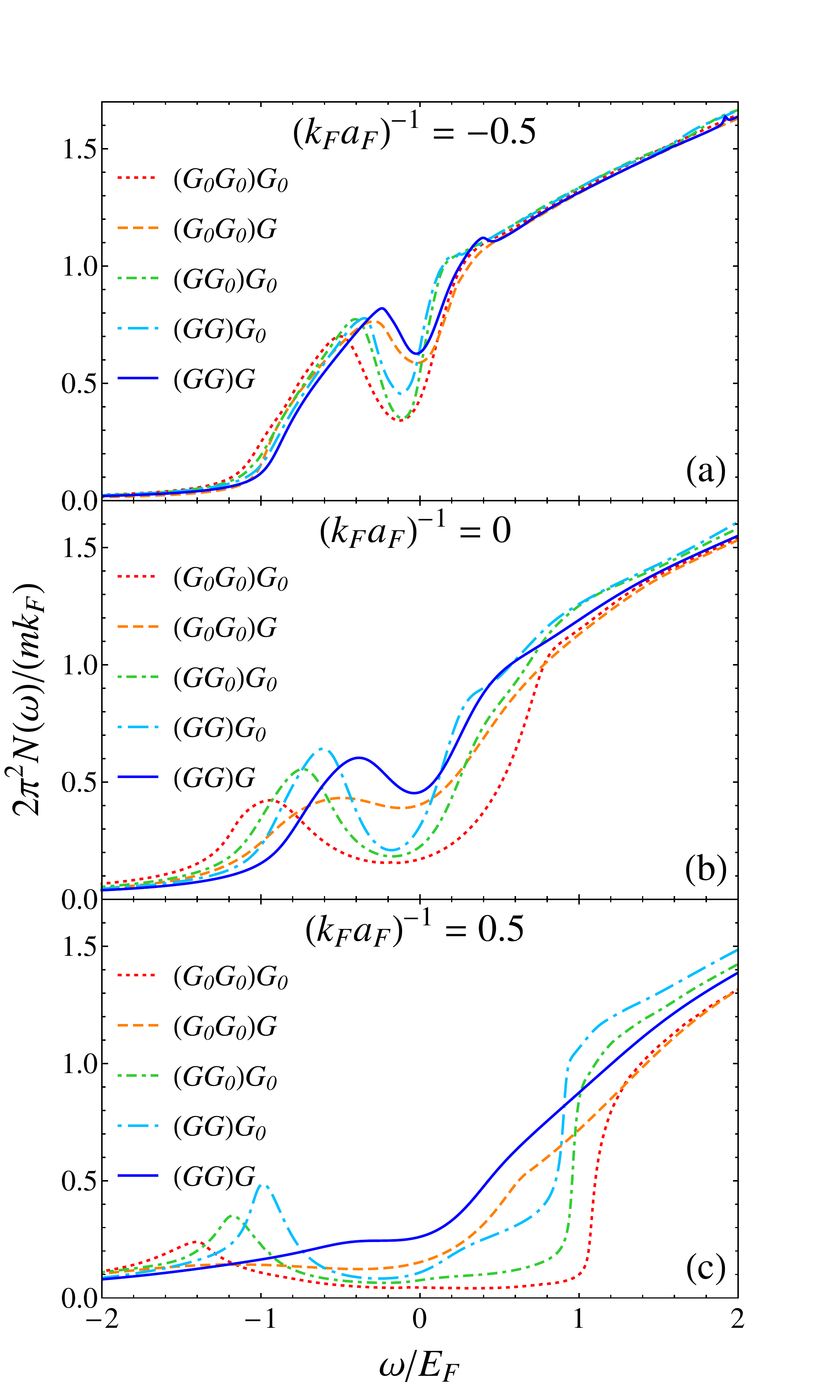}
\caption{(Color online) Density of states $N(\omega)$ at $T=T_c$ for the coupling values $(k_{F} a_{F})^{-1}=(-0.5,0.0,0.5)$ within alternative $t$-matrix approaches. 
                                    The non-interacting density of states $N_{0}=m k_{F}/(2\pi^2)$ per spin component at the Fermi level is used to normalize $N(\omega)$.}
\label{Figure-9}
\end{center}
\end{figure} 

\vspace{0.05cm}
\begin{center}
{\bf C. Single-particle density of states}
\end{center}

To avoid reference to a specific wave vector in the the single-particle spectral function, yet maintaining the main features of its frequency dependence, one can integrate $A(\mathbf{k},\omega)$ over all $\mathbf{k}$ and obtain the \emph{density of states}:
\begin{equation}
N(\omega) = \int \!\! \frac{d \mathbf{k}}{(2\pi)^3} \, A(\mathbf{k},\omega) .
\label{equation-Nw}
\end{equation}
Figure~\ref{Figure-9} shows the density of states obtained in this way at $T_{c}$ for the coupling values $(k_{F}a_{F})^{-1} = (-0.5,0.0,0.5)$, within the alternative $t$-matrix approaches. 
In all cases, a depletion is apparent in the density of states about $\omega=0$. 
The energy width of this depletion is associated to a \emph{pseudo-gap} that develops in the normal phase above $T_{c}$ due to pairing fluctuations, as a precursor of the pairing gap that occurs in the superfluid phase 
below $T_{c}$. 
Although all $t$-matrix approaches present evidence of a pseudo-gap, its detailed structure depends on the specific approach. 
Let's consider, for instance, panel (b) of Fig.~\ref{Figure-9} for the coupling $(k_{F} a_{F})^{-1}=0$.
Here we observe that, similarly to $A(\mathbf{k},\omega)$, also for $N(\omega)$ the approaches can be divided in two classes, namely, those that dress the fermionic propagator 
$G^{(\mathrm{c})}$ in Eq.~(\ref{equation-Sigma_k}) and those that do not.
The first ones present only a narrow and shallow pseudo-gap feature, while the second ones present a wide and deep pseudo-gap feature (which is especially amplified by the $(G_{0} G_{0})G_{0}$ approach).
This is, of course, a direct consequence of the behavior of the single-particle spectral functions shown in Fig.~\ref{Figure-8}, where the single peak in $A(\mathbf{k},\omega)$ that crosses $\omega=0$ 
tends to partially fill the pseudo-gap region for the $(GG)G$ and $(G_{0} G_{0})G$ approaches, while for the $(GG)G_{0}$, $(GG_{0})G_{0}$ and $(G_{0} G_{0})G_{0}$ approaches a depletion persists
about $\omega=0$ owing to the double-peak structure of $A(\mathbf{k},\omega)$. 
For the weaker coupling $(k_{F} a_{F})^{-1}=-0.5$ (cf. panel (a) of Fig.~\ref{Figure-9}), on the other hand, the density of states has the overall shape 
$N_{0}(\omega)= m^{3/2} \sqrt{2(\omega+\mu)}/(2\pi^{2})$ of the non-interacting system, with only a rather minor pseudo-gap feature occurring about $\omega=0$.         
While for this coupling the qualitative behavior of $N(\omega)$ is essentially the same for all approaches, the $(GG)G$ and $(G_{0} G_{0})G$ approaches still have a weaker pseudo-gap behavior than the other approaches. 
Finally, for the coupling $(k_{F} a_{F})^{-1}=0.5$ (cf. panel (c) of Fig.~\ref{Figure-9}) more marked differences appear among the various approaches. 
In particular, the $(GG)G_{0}$, $(GG_{0})G_0$, and $(G_{0} G_{0})G_{0}$ approaches all present quite a wide and deep pseudo-gap feature, the $(G_{0} G_{0})G$ approach presents a wide but rather shallow 
pseudo-gap feature, and the $(GG)G$ approach shows almost no evidence of a pseudo-gap (to the extent that the density of states does not even go through a local minimum near $\omega = 0$).

\begin{figure}[t]
\begin{center}
\includegraphics[width=7.0cm,angle=0]{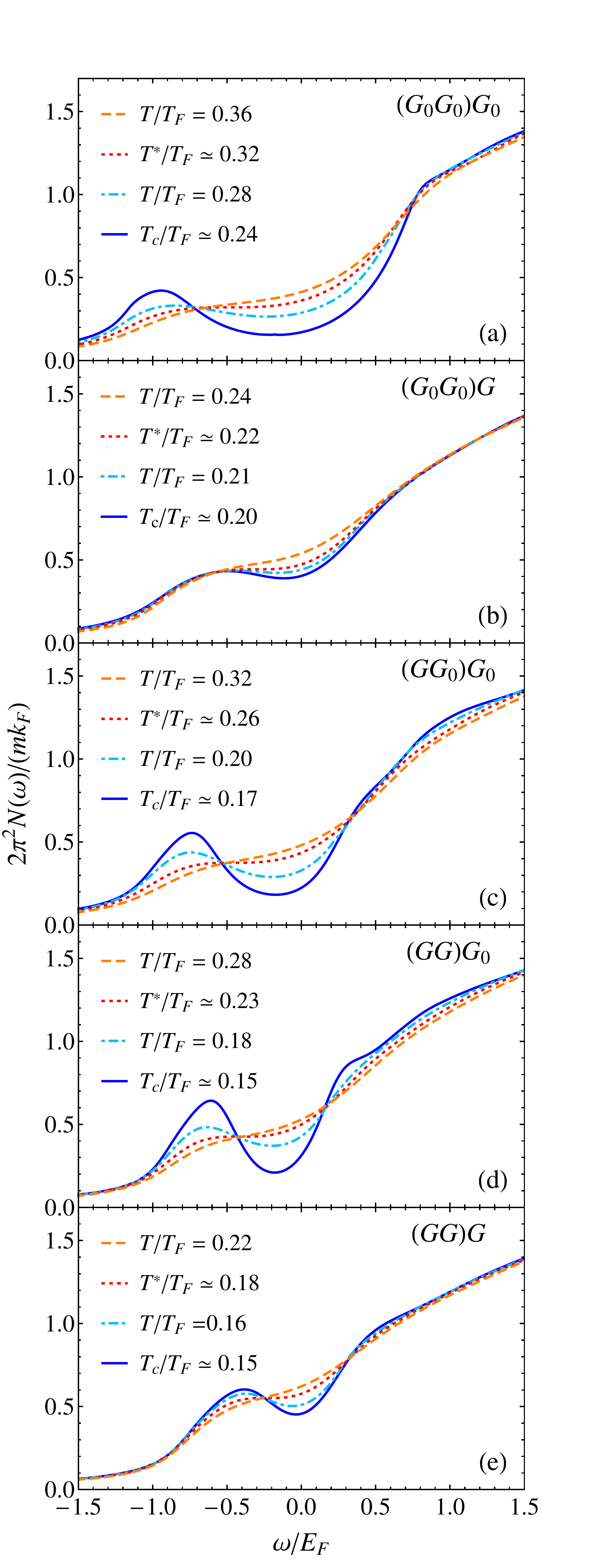}
\caption{(Color online) Evolution of the density of states $N(\omega)$ at unitarity for temperatures $T \ge T_{c}$ within alternative $t$-matrix approaches.}
\label{Figure-10}
\end{center}
\end{figure} 

\vspace{0.05cm}
\begin{center}
{\bf D. Pseudo-gap temperature}
\end{center}

For given coupling, the evolution of the shape of $N(\omega)$ vs $\omega$ can be followed for increasing temperature starting from $T_{c}$.
As an example, Fig.~\ref{Figure-10} shows this temperature evolution when $(k_{F} a_{F})^{-1}=0$ for each $t$-matrix approach we are considering. 
Quite generally, the depletion of the pseudo-gap region about $\omega = 0$ gradually fades away upon increasing temperature, in such a way that a ``crossover'' temperature $T^{*}$ can be identified 
as the highest temperature at which the local minimum of $N(\omega)$ near $\omega = 0$ eventually disappears. 

\begin{figure}[t]
\begin{center}
\includegraphics[width=8.0cm,angle=0]{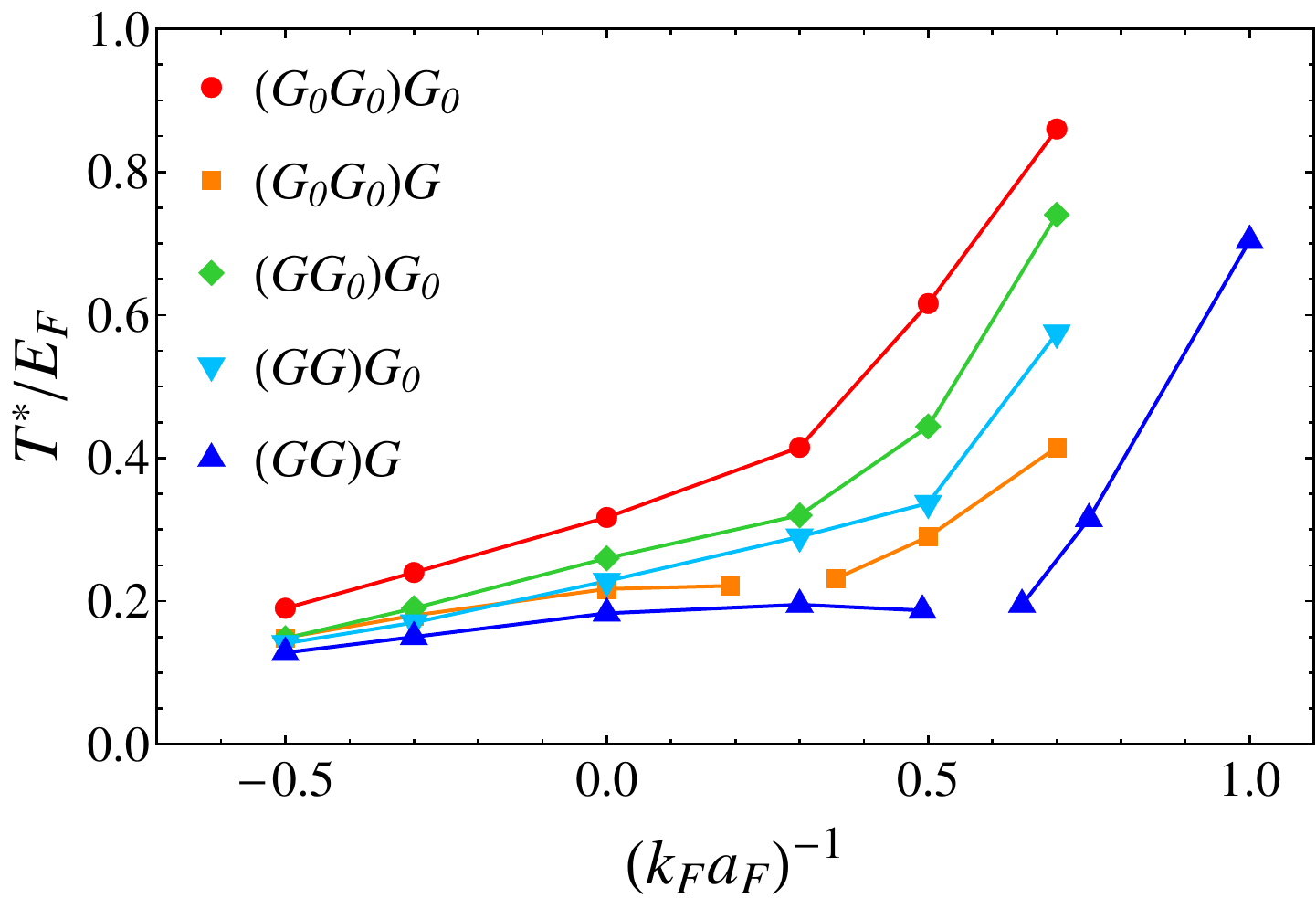}
\caption{(Color online) Temperature $T^{*}$ for the appearance of a pseudo-gap in the density of states, as a function of $(k_{F} a_{F})^{-1}$ within alternative $t$-matrix approaches. 
                                    The non-connected regions that appear for the $(G_{0} G_{0})G$ and $(GG)G$ approaches signal that there $T^{*}$ cannot be identified, to the extent that a local 
                                    minimum near $\omega = 0$ cannot be found in $N(\omega)$ even at $T=T_{c}$.}
\label{Figure-11}
\end{center}
\end{figure} 

The crossover temperature $T^{*}$ obtained in this way throughout the BCS-BEC crossover is reported in Fig.~\ref{Figure-11} for all $t$-matrix approaches. 
It turns out that the coupling dependence of $T^{*}$ differs considerably for the various approaches, especially for positive couplings on the BEC side of unitarity. 
In particular, one notices that $T^{*}$ is considerably suppressed for the $(GG)G$ and $(G_{0} G_{0})G$ approaches with respect to the other approaches. 
In addition, for the $(GG)G$ and $(G_{0} G_{0})G$ approaches there occurs a coupling interval where $T^{*}$ cannot be defined, because the density of states $N(\omega)$ does not have a local minimum 
near $\omega=0$ even at $T=T_{c}$. 
This occurs when $(k_{F} a_{F})^{-1} \simeq 0.3$ for the $(G_{0} G_{0})G$ approach and when $(k_{F} a_{F})^{-1} \simeq 0.6$ for the $(GG)G$ approach. 
Finally, for all approaches $T^{*}$ begins to increase rapidly with coupling around $(k_{F} a_{F})^{-1} \simeq 0.5-0.7$. 
This is because this coupling regime is where the actual \emph{crossover} occurs, from a pseudo-gap phase where the depletion in the density of states is rather shallow and due to a truly \emph{many-body} effect, to a normal-Bose-gas phase where the depletion in the density of states becomes deep and is just evidence of the \emph{two-body} binding energy of composite bosons \cite{Marsiglio-2015}.

\vspace{0.05cm}
\begin{center}
{\bf E. Luttinger wave vector}
\end{center}

The above crossover, between the pseudo-gap and normal-Bose-gas phases, can be characterized in terms of the Luttinger wave vector $k_{L}$.
This wave vector was originally considered in Ref.~\cite{Perali-2011} within the $(G_{0} G_{0})G_{0}$ approach, as the wave vector $k$ at which the back-bending of the lower branch $\epsilon(k)$ 
of the single-particle dispersion occurs. 
In Ref.~\cite{Perali-2011}, this branch was obtained by following the $k$-dependence of the low-energy peak in the single-particle spectral function $A(\bm{k},\omega)$ (cf. panel (a) of 
Fig.~\ref{Figure-8}), and then by fitting the dispersion of the lower branch $\epsilon(k)$ obtained in this way through a BCS-like form
\begin{equation}
\frac{\epsilon(k)}{E_{F}}= \mu' - \sqrt{(k^{2}-k_{L}^{2})^{2} + \Delta'^{2}} .
\label{BCS-lile-dispersion}
\end{equation}
Here, $\mu'$ and $\Delta'$ are fitting parameters (in units of $E_{F}$), with the energy shift $\mu'$ needed to account for the dispersion $\epsilon(k)$ away from the weak-coupling regime, 
and $k$ and $k_{L}$ are in units of $k_{F}$.
On physical grounds, a non-vanishing value of $k_{L}$ signals the presence of an underlying Fermi surface, which endows the system with a persistent fermionic character even in the presence of a strong attractive inter-particle interaction.                              
In these terms, the crossover from the pseudo-gap to the normal-Bose-gas phase is considered complete only when $k_{L}$ reaches zero at some critical coupling, thus signaling the eventual disappearance of the underlying Fermi surface. 

The above definition of $k_{L}$, introduced in Ref.~\cite{Perali-2011} within the $(G_{0} G_{0})G_{0}$ approach, can as well be extended to the $(G G_{0})G_{0}$ and $(GG)G_{0}$ approaches here considered, 
whereby a lower branch of the dispersion can be clearly identified from the spectra of $A(\bm{k},\omega)$ (cf. panels (c) and (d) of Fig.~\ref{Figure-8}) and a back-bending occurs.  
However, this definition of $k_{L}$ cannot be transferred to the $(G_{0} G_{0})G$ and $(GG)G$ approaches, where only a single peak appears in the $A(\bm{k},\omega)$ (cf. panels (b) and (e) of Fig.~\ref{Figure-8}) and there is no observable back-bending in the dispersion for most couplings. 
As a consequence, for the latter two approaches we make use of an alternative operative definition of $k_{L}$ which is closer in spirit to the description one would adopt for a Fermi liquid, 
and identify $k_{L}$ as the wave vector for which the single peak in $A(\mathbf{k},\omega)$ passes through $\omega=0$. 
Nevertheless, this definition appears to work properly up to when $k_{L}/k_{F} \gtrsim 0.5$, because at that point a double-peak structure begins to appear 
in $A(\mathbf{k},\omega)$ even for the $(G_{0} G_{0})G$ and $(GG)G$ approaches. 
To be able to extend the $k_{L}$-vs-coupling curve up to $k_{L} \rightarrow 0$ when the full collapse of the underlying Fermi surface occurs, for the coupling regime where the double peak occurs we thus found it more appropriate to identify $k_{L}$ as the wave vector at which the peaks in $A(\mathbf{k},\omega)$ at negative and positive frequencies mutually exchange the height of their maxima.

\begin{figure}[t]
\begin{center}
\includegraphics[width=8.0cm,angle=0]{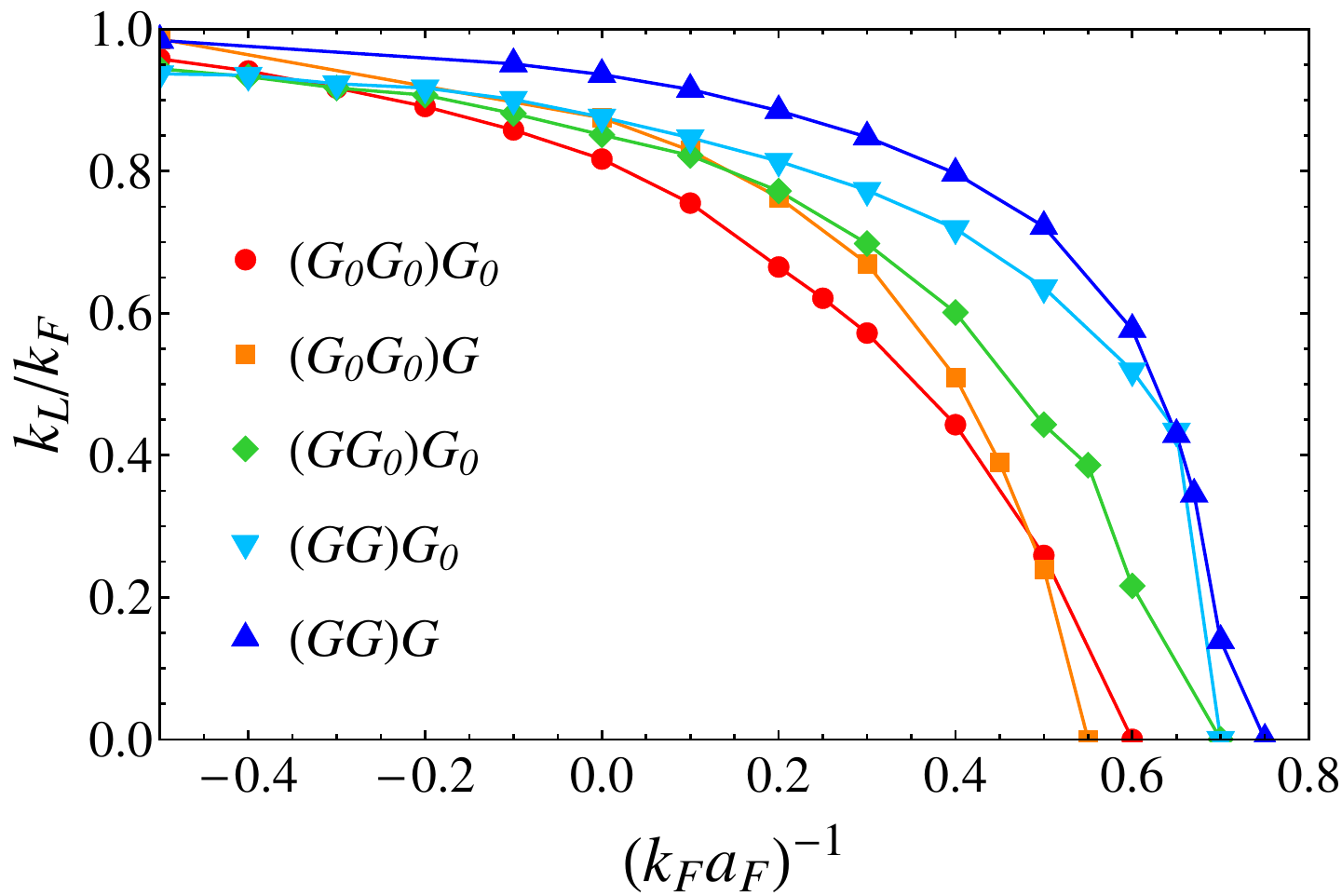}
\caption{(Color online) Luttinger wave vector $k_{L}$ (in units of $k_{F}$) vs $(k_{F} a_{F})^{-1}$ obtained at $T=T_{c}$ within alternative $t$-matrix approaches 
                                     (the definition of $k_{L}$ for the different approaches is given in the text).}
\label{Figure-12}
\end{center}
\end{figure} 

Figure~\ref{Figure-12} reports the values of the Luttinger wave vector $k_{L}$ as a function of $(k_{F} a_{F})^{-1}$ obtained for all $t$-matrix approaches using the procedures described above. 
Here, depending on the approach, $k_{L}$ is seen to vanish for couplings values in the rather narrow range between $0.55$ and $0.7$, with the largest critical coupling reached by the $(GG)G$ fully self-consistent 
approach.

\section{Concluding remarks and perspectives} 
\label{sec:conclusions}

In this paper, we have performed a systematic theoretical study about several variants of the $t$-matrix approximation for an attractive dilute Fermi gas in its normal phase above the superfluid critical temperature $T_{c}$.
This study was extended to the whole BCS-BEC crossover and has regarded both thermodynamic and dynamical quantities that characterize the Fermi gas.
Although these variants of the $t$-matrix approximation have already been separately considered in the literature, the novelty here is that all these variants have been treated on equal footing (with the same numerical accuracy also having been pursued for all of them) and in an unbiased way, in order to evidence their individual virtues and shortcoming.

As far as the thermodynamic quantities that we have considered are concerned, from one $t$-matrix approach to the other we have found mostly quantitative differences but similar qualitative trends,
apart from the presence vs absence of a maximum for the critical temperature in the intermediate-coupling regime (cf. Fig.~\ref{Figure-3}) and of a residual bosonic interaction affecting the chemical potential in the strong-coupling (BEC) regime (cf. Fig.~\ref{Figure-4}).
The most distinctive differences (not only at a quantitative but also at a qualitative level) among the outcomes of the various $t$-matrix approaches have instead been found for the dynamical quantities, specifically, about the occurrence of a one-peak vs two-peak structure in the single-particle spectral function (cf. Fig.~\ref{Figure-8}).
This qualitative difference appears relevant, not because it hinges on a dispute about the Fermi liquid vs non-Fermi liquid behavior of an attractive Fermi gas at unitarity \cite{Salomon-2011}, but rather because it affects the width (if not the presence itself) of a temperature interval above $T_{c}$ where a pseudo-gap regime would show up, with the simultaneous 
presence of preformed pairs and of an underlying Fermi surface.

That a fully self-consistent diagrammatic approximation may end up in giving (even considerably) smaller values for the \emph{excitation} energies with respect to its non-self-consistent version(s) 
(and possibly ``overshoot the mark'' when comparing with experimental values) has also been evidenced in other physical contexts.
One can specifically refer to the spectra associated with electronic excitations that can be described in terms of the GW approximation, not only for semiconductors and insulators \cite{Kresse-2018}
but also for complex molecules \cite{Rinke-2012}.
In these cases, attempts have recently been made to mitigate the effects of self-consistency by introducing vertex corrections on the GW calculations 
\cite{Kutepov-2018,Berkelbach-2018,footnote-new}.

In the present context of the $t$-matrix approximation for an attractive Fermi gas, too, vertex corrections have recently been included on top of a partially self-consistent version of this approximation, ending up
with rather good results for the critical temperature $T_{c}$ \cite{Pisani-2018-I} and the pairing gap at zero temperature \cite{Pisani-2018-II} throughout the BCS-BEC crossover, when compared with available quantum Monte Carlo calculations and experimental data.
It could therefore be interesting to assess whether, including a similar kind of vertex corrections on top of the fully self-consistent $t$-matrix approximation, may result in more favourable conditions for the presence of a pseudo-gap regime about unitarity.


\vspace{0.2cm}
\begin{center}
\begin{small}
{\bf ACKNOWLEDGMENTS}
\end{small}
\end{center}
\vspace{-0.1cm}

M.P. acknowledges L. Fallani for support and discussions.
This work was partially supported by the Italian MIUR under Contract PRIN-2015 No. 2015C5SEJJ001.


\appendix   
\section{DETAILS OF THE NUMERICAL PROCEDURES FOR ACHIEVING \\ (PARTIAL OR FULL) SELF-CONSISTENCY}
\label{sec:appendix-A}
\vspace{-0.2cm}

In this Appendix, we present in detail the numerical procedures that are needed to implement the cycles of self-consistency depicted schematically in Fig.~\ref{Figure-2}.
For the sake of definiteness, we will specifically consider the fully self-consistent $(GG)G$ approach whose cycle is shown in Fig.~\ref{Figure-2}(d), since all procedures discussed for this approach can as well 
be applied to the partially self-consistent approaches. 
We remark that the procedures here presented for the $(GG)G$ approach are in line with those previously suggested in Ref.~\cite{Haussmann-1994} and partially with those reported in Ref.~\cite{VanHoucke-2013}.

All expressions reported in this Appendix are given in dimensionless units, such that energies are in units of the Fermi energy $E_{F}=k_{F}^{2}/(2m)$ and wave vectors  in units of the Fermi wave vector $k_{F}$. 
Accordingly, the single-particle fermionic propagator $G(\mathbf{k},\omega_{n})$ is in units of $E_{F}^{-1}$, the fermionic self-energy $\Sigma(\mathbf{k},\omega_n)$ in units of $E_{F}$, and the 
particle-particle propagator $\Gamma(\mathbf{Q},\Omega_{\nu})$ in units of $(m k_{F})^{-1}$.
In addition, to further shorten the notation, here we use the symbol $v=(k_{F} a_{F})^{-1}$ for the coupling. 

\vspace{0.05cm}
\begin{center}
{\bf 1. Transforming from $G(\mathbf{k},\omega_{n})$ to $G(\mathbf{r},\tau)$}
\end{center}

The first function to be Fourier transformed in the self-consistent cycle of Fig.~\ref{Figure-2}(d) is the single-particle fermionic propagator $G$. 
For this function the Fourier transform can be done in two steps, namely,
\begin{equation}
G(\mathbf{k},\omega_{n}) \rightarrow G(\mathbf{k},\tau) \rightarrow G(\mathbf{r},\tau)
\label{equation-G_FT_scheme}
\end{equation}
\noindent
with the Fourier transform over the wave vector $\mathbf{k}$ following that over the frequency $\omega_{n}$. 

In the first step, to get the Fourier transform over the frequency $\omega_{n}$, we note that the dressed fermionic propagator $G(\mathbf{k},\omega_{n})$ for large frequencies behaves like the free propagator 
$G_{0}(\mathbf{k},\omega_{n})$:
\begin{equation}
G(\mathbf{k},\omega_{n}) \underset{ \omega_{n} \to \infty}{\simeq} G_{0}(\mathbf{k},\omega_{n}) = \frac{1}{i \omega_{n} - \xi_{\mathbf{k}}}
\end{equation}
where $\xi_{\mathbf{k}} = \mathbf{k}^2-\mu$. 
One can thus calculate numerically the Fourier transform from $\omega_{n}$ to $\tau$ of the difference
\begin{equation}
\Delta G(\mathbf{k},\omega_{n}) = G(\mathbf{k},\omega_{n}) - G_{0}(\mathbf{k},\omega_{n}) ,
\label{difference-G}
\end{equation}
which is easier to obtain than the Fourier transform of $G(\mathbf{k},\omega_{n})$ since the function (\ref{difference-G}) converges like $\sim\omega_{n}^{-5/2}$ for large frequencies.
One then adds to this result the Fourier transform of $G_{0}(\mathbf{k},\omega_{n})$, which is known analytically in the form \cite{FW-1971}
\begin{equation}
G_{0}(\mathbf{k},\tau) = e^{-\xi_{\mathbf{k}}\tau} \big(f(\xi_{\mathbf{k}})-1 \big) , 
\label{equation-G0_k_tau}
\end{equation}
where $f(\xi_{\mathbf{k}}) = ( e^{\beta \xi_{\mathbf{k}}} + 1 )^{-1}$ is the Fermi distribution function.

In the second step, to get the Fourier transform over the wave vector $\mathbf{k}$, it is further convenient to split the expression (\ref{equation-G0_k_tau}) in two parts:
\begin{equation}
G_{0}(\mathbf{k},\tau) = G^{(\mathrm{n})}_{0}(\mathbf{k},\tau) +G^{(\mathrm{a})}_{0}(\mathbf{k},\tau)
\label{equation-G0n+a}
\end{equation}
where
\begin{eqnarray}
G^{(\mathrm{n})}_{0}(\mathbf{k},\tau) & = & e^{-\xi_{\mathbf{k}}\tau} f(\xi_{\mathbf{k}}) 
\label{equation-G0n} \\
G^{(\mathrm{a})}_{0}(\mathbf{k},\tau) & = & - e^{-\xi_{\mathbf{k}}\tau}.
\label{equation-G0a} 
\end{eqnarray}
Here, the labels $(\mathrm{n})$ and $(\mathrm{a})$ signify that the term (\ref{equation-G0n}) has to be numerically Fourier transformed over $\mathbf{k}$, whereas the term (\ref{equation-G0a}) 
admits an .pdfr transform of the form:
\begin{equation}
G^{(\mathrm{a})}_{0} (\mathbf{r},\tau) = - \frac{e^{\mu \tau}e^{-\frac{r^2}{4 \tau}}}{8 \pi^{3/2} \tau^{3/2}} 
\label{equation-G0a_r}
\end{equation}
which in the limit $\tau \rightarrow 0^{+}$ is a representation of the Dirac delta function $\delta^{3}(\mathbf{r})$. 
This property is related to the anticommutator between fermionic field operators that appears in the fermionic propagator when passing from $\tau = 0^{-}$ to $\tau = 0^{+}$. 
The term (\ref{equation-G0a_r}) thus describes the singular behavior when $(\mathbf{r},\tau) \to 0^{+}$, not only for the free propagator $G_{0}$ but also for the dressed propagator $G$. 
It is then convenient to define a new function
\begin{equation}
\tilde{\Delta}G(\mathbf{k},\tau) = G(\mathbf{k},\tau) - G^{(\mathrm{a})}_{0}(\mathbf{k},\tau) ,
\label{new-difference-G}
\end{equation}
which can be readily Fourier transformed over $\mathbf{k}$ numerically. 
The desired Fourier transform $G(\mathbf{r},\tau)$ is eventually obtained by adding $G^{(a)}_{0}(\mathbf{r},\tau)$ of Eq.~(\ref{equation-G0a_r}) to the Fourier transform of 
$\tilde{\Delta}G(\mathbf{k},\tau)$ \cite{footnote-2}.

\vspace{0.05cm}
\begin{center}
{\bf 2. Transforming from $\Gamma(\mathbf{Q},\Omega_{\nu})$ to $\Gamma(\mathbf{r},\tau)$}
\end{center}

The next function to be Fourier transformed in the self-consistent cycle of Fig.~\ref{Figure-2}(d) is the particle-particle propagator $\Gamma$.
Also in this case, the Fourier transform can be done in two steps, namely,
\begin{equation}
\Gamma(\mathbf{Q},\Omega_{\nu}) \rightarrow  \Gamma(\mathbf{Q},\tau) \rightarrow  \Gamma(\mathbf{r},\tau),
\label{equation-Gamma_FT_scheme}
\end{equation}
with the Fourier transform over the wave vector $\mathbf{Q}$ following again that over the frequency $\Omega_{\nu}$.

To perform the first Fourier transform over $\Omega_{\nu}$, we begin by noting that the large-frequency behavior of the particle-particle propagator $\Gamma(\mathbf{Q},\Omega_{\nu})$ coincides with that of its
non-self-consistent counterpart taken in the strong-coupling limit $- \beta \mu \gg 1$, which is given by the expression \cite{Pieri-2000}:
\begin{eqnarray}
\Gamma(\mathbf{Q},\Omega_{\nu})  & \underset{ \Omega_{\nu} \rightarrow \infty}{\simeq} & \Gamma_{\mathrm{sc}}(\mathbf{Q},\Omega_{\nu}) 
\nonumber \\
& = & - \, \frac{4 \pi}{v-\sqrt{\frac{Q^{2}}{4}-\mu-i\frac{\Omega_{\nu}}{2}}} .
\label{equation-Gamma_sc_Q_Omega}
\end{eqnarray}
However, using $\Gamma_{\mathrm{sc}} (\mathbf{Q}, \Omega_{\nu})$ as the reference function to be subtracted in the Fourier transform may lead to problems, because for $\Omega_{\nu}=0$ the function 
(\ref{equation-Gamma_sc_Q_Omega}) has a pole at $|\mathbf{Q}|=2\sqrt{v^{2} + \mu}$ when $v>0$. 
To the extent that we are interested only in taking care of the large-frequency behavior of $\Gamma(\mathbf{Q},\Omega_{\nu})$, we are led to introduce the new reference function
\begin{equation}
\Gamma_{\mathrm{sc}}' (\mathbf{Q}, \Omega_{\nu}) = \Gamma_{\mathrm{sc}} (\mathbf{Q}, \Omega_{\nu}) - \Gamma_{\mathrm{sc}} (\mathbf{Q}, \Omega_{\nu=0})
\label{equation-Gamma_sc-prime-_Q_Omega}
\end{equation}
with the zero-frequency term removed from the expression (\ref{equation-Gamma_sc_Q_Omega}).
Although the Fourier transform of the function (\ref{equation-Gamma_sc-prime-_Q_Omega}) cannot be calculated analytically, it can be computed with limited effort by writing it as an integral over the complex $z$-plane as follows:
\begin{eqnarray}
\Gamma_{\mathrm{sc}}'(\mathbf{Q},\tau) & = & T \sum_{\nu} e^{-i \Omega_{\nu} \tau} \Gamma_{\mathrm{sc}}'(\mathbf{Q},\Omega_{\nu}) 
\nonumber \\
& = & \frac{1}{2 \pi i} \oint_{\mathcal{C}} dz \frac{e^{z \tau}}{(e^{\beta z} - 1)} \Gamma_{\mathrm{sc}}'(\mathbf{Q},z)
\label{integral-over-complex-plane}
\end{eqnarray}
where the contour $\mathcal{C}$ surrounds the poles of the Bose function $b(z)=1/(e^{\beta z} - 1)$ on the imaginary axis. 
Here, the function $\Gamma_{\mathrm{sc}}'(\mathbf{Q},z)$ has a branch cut along the negative real axis starting at $z_{\mathrm{c}} = 2(\mu-\mathbf{Q}^2/4)$ as well as a pole at 
$z_{\mathrm{p}} =2 v^{2}+z_c$ when $v>0$.
The integral in Eq.~(\ref{integral-over-complex-plane}) then reduces to the calculation of an integral along the branch cut and of the residue of the pole, and can accordingly be split in the following way: 
\begin{eqnarray}
\Gamma'_{\mathrm{sc}}(\mathbf{Q},\tau) & = & {\Gamma}_{\mathrm{sc}}^{(\mathrm{n})}(\mathbf{Q},\tau) + \Gamma_{\mathrm{sc}}^{(\mathrm{a})}(\mathbf{Q},\tau) 
\label{equation-Gamma'_sc_Q_tau} \\
& + & \Gamma_\mathrm{res} (\mathbf{Q},\tau)  -T \, \mathrm{Re}[ \Gamma_{\mathrm{sc}}(\mathbf{Q},\Omega_{\nu=0)}] 
\nonumber
\end{eqnarray}
where the first two terms are contributed by the branch cut and the third term by the pole.
Like in Eq.~(\ref{equation-G0n+a}), the labels $(\mathrm{n})$ and $(\mathrm{a})$ in the first line of Eq.~(\ref{equation-Gamma'_sc_Q_tau}) signify that these contributions are calculated numerically or analytically, respectively.

The first term in Eq.~(\ref{equation-Gamma'_sc_Q_tau}) can be cast in the form
\begin{equation}
\Gamma_{\mathrm{sc}}^{(\mathrm{n})}(\mathbf{Q},\tau)= \frac{8 \sqrt{2} e^{z_{\mathrm{c}} \tau}}{\sqrt{\tau}} \!\! \int_0^{+\infty} \!\!\!\!\!\! dx \frac{ e^{-x^2}\, x^2}{(e^{-\beta(z_{\mathrm{c}}-x^2/\tau)}-1)(x^2+2 \tau v^{2})}
\end{equation}
for $z_{\mathrm{c}}<0$ (that is, for $\mu<\mathbf{Q}^2/4$), and in the form
\begin{widetext} 
\begin{eqnarray}
\Gamma_{\mathrm{sc}}^{(\mathrm{n})}(\bm{Q},\tau) & = & \frac{8 \sqrt{2} e^{z_{\mathrm{c}} \tau}}{\sqrt{\tau}} \int_{\sqrt{\tau(z_{\mathrm{c}}-z_0)}}^{+\infty} dx \frac{ e^{-x^2}\, x^2}{(e^{-\beta(z_c-x^2/\tau)}-1)(x^2+2 \tau 
v^{2})}
\\
& + & 8 \sqrt{\frac{2}{\tau}} \int_{0}^{\sqrt{\tau(z_{\mathrm{c}}-z_0)}} dx \frac{1}{e^{-\beta(z_{\mathrm{c}}-x^2/\tau)}-1} \bigg( \frac{e^{z_{\mathrm{c}}\tau-x^2} x^2}{x^2+2\tau v^{2}} - \frac{x \sqrt{z_{\mathrm{c}}}}{z_{\mathrm{p}} \sqrt{\tau}} \bigg) +\frac{4 \sqrt{2 z_{\mathrm{c}}}}{z_{\mathrm{p}}} \bigg( z_{0}-z_{\mathrm{c}}-T \, \ln \bigg| \frac{e^{-\beta z_{\mathrm{c}}}-1}{e^{-\beta z_{0}}-1} \bigg| \bigg)
\nonumber
\end{eqnarray}
\end{widetext}
for $z_{\mathrm{c}} \ge 0$ (that is, for $\mu \ge \mathbf{Q}^2/4$), where $z_{0}$ is any point along the negative real axis (typically, we have taken $z_{0} = - 2T$).
The second term in Eq.~(\ref{equation-Gamma'_sc_Q_tau}) has  instead the semi-analytic form:
\begin{equation}
\Gamma_{\mathrm{sc}}^{(\mathrm{a})}(\mathbf{Q},\tau) = 4 \sqrt{2 \pi} \,\, c(\tau,v) \, \frac{e^{2\mu \tau} e^{-\mathbf{Q}^{2}\tau/2}}{\sqrt{\tau}}
\label{equation-Gamma0a_Q_tau}
\end{equation}
with the coefficient $c(\tau,v)$ given by
\begin{equation}
c(\tau,v) = \frac{2}{\sqrt{\pi}} \int_{0}^{+\infty} dx \, \frac{x^2 \, e^{-x^2} }{x^2+2\tau v^{2}} \, .
\label{equation-c_Gamma0a}
\end{equation}
Note that this coefficient is unity for $\tau \to 0^{+}$ or $v=0$. 

The term (\ref{equation-Gamma0a_Q_tau}) admits also an .pdfr transform from $(\mathbf{Q},\tau)$ to $(\mathbf{r},\tau)$, which is given by:
\begin{equation}
\Gamma_{\mathrm{sc}}^{(\mathrm{a})}(\mathbf{r},\tau) = \frac{2 \, c(\tau,v) \, e^{2\mu \tau} e^{-\frac{\mathbf{r}^2}{2\tau}}}{\pi\tau^2} \, .
\label{equation-Gamma0a_r_tau}
\end{equation}
One can show that this is the leading term of the singular behavior of the full $\Gamma(\mathbf{r},\tau)$ for $(\mathbf{r},\tau) \rightarrow 0^{+}$, not only in the strong-coupling regime but also 
throughout the BCS-BEC crossover. 
Finally, the third term in Eq.~(\ref{equation-Gamma'_sc_Q_tau}) is the residue of the pole at $z=z_{\mathrm{p}}$, given by:
\begin{equation}
\Gamma_\mathrm{res}(\mathbf{Q},\tau)=-\theta (v) \frac{16 \pi v \, e^{z_{\mathrm{p}} \tau}}{e^{\beta z_{\mathrm{p}}} -1} \, .
\end{equation}
One can also show that the divergence of this term for $z_{\mathrm{p}} = 0$ (that is, for $|\mathbf{Q}|=2\sqrt{v^{2} + \mu}$) is exactly compensated by the fourth term of (\ref{equation-Gamma'_sc_Q_tau}),
in such a way that $\Gamma'_{\mathrm{sc}} (\mathbf{Q},\tau)$ is always a smooth function of $\mathbf{Q}$. 

At this point, we calculate numerically the Fourier transform from $(\mathbf{Q},\Omega_{\nu})$ to $(\mathbf{Q},\tau)$ of the difference
\begin{equation}
\Delta \Gamma(\mathbf{Q},\Omega_{\nu})= \Gamma(\mathbf{Q},\Omega_{\nu})-\Gamma'_{\mathrm{sc}}(\mathbf{Q},\Omega_{\nu})
\label{difference-Gamma}
\end{equation}
to obtain $\Delta \Gamma(\mathbf{Q},\tau)$, and then add to it $\Gamma'_{\mathrm{sc}}(\mathbf{Q},\tau)$ given by Eq.~(\ref{equation-Gamma'_sc_Q_tau}) to obtain $\Gamma(\mathbf{Q},\tau)$.

Finally, for the remaining Fourier transform from $(\mathbf{Q},\tau)$ to $(\mathbf{r},\tau)$, we can make use of $\Gamma_{\mathrm{sc}}^{(\mathrm{a})}(\mathbf{Q},\tau)$ defined by
Eqs.~(\ref{equation-Gamma0a_Q_tau}) and (\ref{equation-c_Gamma0a}) to subtract the leading term of the singular behavior when $(\mathbf{r},\tau) \to 0^{+}$, similarly to what we did in Eq.~(\ref{new-difference-G}).
Accordingly, we define the difference function:
\begin{equation}
\tilde{\Delta} \Gamma(\mathbf{Q},\tau) = \Gamma(\mathbf{Q},\tau) - \Gamma_{\mathrm{sc}}^{(\mathrm{a})}(\mathbf{Q},\tau)
\end{equation}
and Fourier transform it from $\mathbf{Q}$ to $\mathbf{r}$ to obtain $\tilde{\Delta} \Gamma(\mathbf{r},\tau)$. 
The desired function $\Gamma(\mathbf{r},\tau)$ eventually results by adding to $\tilde{\Delta} \Gamma(\mathbf{r},\tau)$ the semi-analytic expression (\ref{equation-Gamma0a_r_tau}) for 
$\Gamma_{\mathrm{sc}}^{(\mathrm{a})}(\mathbf{r},\tau)$ \cite{footnote-3}.

\vspace{0.05cm}
\begin{center}
{\bf 3. Transforming from $\Sigma(\mathbf{r},\tau)$ to $\Sigma(\mathbf{k},\omega_{n})$}
\end{center}

The last function to be Fourier transformed in the self-consistent cycle of Fig.~\ref{Figure-2}(d) is the self-energy $\Sigma$.
Also in this case, the Fourier transform is done in two steps, namely,
\begin{equation}
\Sigma(\mathbf{r},\tau) \rightarrow \Sigma(\mathbf{k},\tau) \rightarrow \Sigma(\mathbf{k},\omega_{n})
\label{equation-Sigma_FT_scheme}
\end{equation}
where now one first transforms from $\mathbf{r}$ to $\mathbf{k}$ and then from $\tau$ to $\omega_{n}$, in the reversed direction to what was done for $G$ (cf. Eq.~(\ref{equation-G_FT_scheme})) 
and for $\Gamma$ (cf. Eq.~(\ref{equation-Gamma_FT_scheme})).
This is because from Eq.~(\ref{equation-Sigma_r}), once $G(\mathbf{r},\tau)$ and $\Gamma(\mathbf{r},\tau)$ are know, $\Sigma(\mathbf{r},\tau)$ is also known. 

From Eq.~(\ref{equation-Sigma_r}) one can also determine the singular behavior of $\Sigma(\mathbf{r},\tau)$ for $(\mathbf{r},\tau) \rightarrow 0^{+}$ in terms of those of $G(\mathbf{r},\tau)$ 
[cf. the discussion after Eq.~(\ref{equation-G0a_r})] and of $\Gamma(\mathbf{r},\tau)$ [cf.~the discussion after Eq.~(\ref{equation-Gamma0a_r_tau})].
To this end, we rewrite Eq.~(\ref{equation-Sigma_r}) in the alternative forms:
\begin{eqnarray}
\Sigma(\mathbf{r},\tau) & = & -2 \, \Gamma(\mathbf{r},\tau) \, G(-\mathbf{r},-\tau)  
\label{inverse-convolution-up} \\
& = & 2 \, \Gamma(\mathbf{r},\tau) \, G(\mathbf{r},\beta-\tau) 
\label{inverse-convolution-down}
\end{eqnarray}
(where the factor of two originates by having expressed the particle-particle propagator $\Gamma$ in terms of dimensionless quantities).   
In the second line we have used the spatial isotropy and the anti-periodicity in $\tau$ of the fermionic propagator, to write \mbox{$G(-\mathbf{r},-\tau)=-G(\mathbf{r},\beta-\tau)$}. 
As discussed previously, both $G(\mathbf{r},\tau)$ and $\Gamma(\mathbf{r},\tau)$ are strongly peaked for $(\mathbf{r},\tau) \rightarrow 0^{+}$. 
We then expect the singular behavior of $\Sigma(\mathbf{r},\tau)$ to be captured by the following alternative expressions:
\begin{equation}
\Sigma^{(+)}(\mathbf{r},\tau) \simeq 2 \, \Gamma(\mathbf{r},\tau) \, G(\mathbf{r}=0,\beta^{-})
\label{equation-Sigma+_r_tau}
\end{equation}
in the limit $(\mathbf{r},\tau) \rightarrow (0,0^{+})$ and
\begin{equation}
\Sigma^{(-)}(\mathbf{r},\tau) \simeq 2 \, \Gamma(\mathbf{r}=0,\beta^{-}) \, G(\mathbf{r},\beta-\tau)
\label{equation-Sigma-_r_tau}
\end{equation}
in the limit $(\mathbf{r},\tau) \rightarrow (0,\beta^{-})$. 
Out of the above terms, (\ref{equation-Sigma+_r_tau}) is the dominant one because $\Gamma(\mathbf{r},\tau)$ [cf. Eq.~(\ref{equation-Gamma0a_r_tau})] is more strongly peaked than $G(\mathbf{r},\tau)$ [cf. Eq.~(\ref{equation-G0a_r})] in the limit $\tau \rightarrow 0^{+}$.  
 
Accordingly, in order to perform the Fourier transform of $\Sigma(\mathbf{r},\tau)$ from $\mathbf{r}$ to $\mathbf{k}$, we consider the difference
\begin{equation}
\Delta \Sigma(\mathbf{r},\tau) = \Sigma(\mathbf{r},\tau) - \Sigma^{(+)} (\mathbf{r},\tau)
\label{difference-Sigma}
\end{equation}
with the term (\ref{equation-Sigma+_r_tau}) only. 
In addition, to the extent that in Eq.~(\ref{difference-Sigma}) we are interested in the leading behavior of $\Sigma^{(+)}$ for $(\mathbf{r},\tau) \rightarrow (0,0^{+})$, we can approximate the particle-particle propagator 
$\Gamma(\mathbf{r},\tau)$ in Eq.~(\ref{equation-Sigma+_r_tau}) by the analytic result (\ref{equation-Gamma0a_r_tau}) and write
\begin{equation}
\Sigma^{(+)}(\mathbf{r},\tau) = - 2 n \, \frac{\, e^{-\frac{r^2}{2\tau}}}{ \pi \tau^2} \, ,
\label{equation-Sigma+_ref_r_tau}
\end{equation}
where we have consistently set $c(\tau,v) \to 1$ and $e^{\mu \tau} \rightarrow 1$ and used the result $n = -2 \, G(\mathbf{r}=0,\beta^{-})$ in terms of the fermionic density $n$ (in dimensionless units). 
[For the $t$-matrix approaches that use an external $G_{0}$ in the place of $G$, the expression (\ref{equation-Sigma+_ref_r_tau}) needs to be modified by replacing $n$ by the free-fermion density 
$n_{0} = -2 \, G_{0}(\mathbf{r}=0,\beta^{-})$.]

The expression (\ref{equation-Sigma+_ref_r_tau}) has the further advantage that its Fourier transform from $\tau$ to $\omega_{n}$ can be obtained analytically in terms of the error function
\cite{AS-1972}, in the form:
\begin{equation}
\Sigma^{(+)}(\mathbf{k},\omega_{n}) = - 8 \pi n \,\, \frac{\mathrm{erf}\big(\sqrt{\beta (\mathbf{k}^{2}-2 i \omega_{n})/2} \, \big)}{\sqrt{\mathbf{k}^{2}-2 i \omega_{n}}} \, .
\label{equation-Sigma+_ref_k_omega}
\end{equation}
This expression correctly reproduces the leading $\omega_{n}^{-1/2}$ behavior of $\Sigma(\mathbf{k},\omega_{n})$ for large frequencies. 
[Similarly, $\Sigma^{(-)}$ of Eq.~(\ref{equation-Sigma-_r_tau}) can be shown to behave like $\sim \omega_{n}^{-1}$ for large frequencies \cite{VanHoucke-2013}, thereby confirming that this is a sub-leading contribution also in the frequency domain.] 

Once the difference (\ref{difference-Sigma}) has been Fourier transformed numerically (first from $\mathbf{r}$ to $\mathbf{k}$ and then from $\tau$ to $\omega_{n}$) according to the above prescriptions to obtain 
$\Delta \Sigma(\mathbf{k},\omega_{n})$, one can eventually add to it the analytic expression (\ref{equation-Sigma+_ref_k_omega}) and obtain the desired function $\Sigma(\mathbf{k},\omega_{n})$.

\section{OPTIMIZING THE CONVERGENCE TOWARDS SELF-CONSISTENCY}
\label{sec:appendix-B}
\vspace{-0.2cm}

In this Appendix, we discuss the procedures that we have adopted to achieve optimal convergence toward self-consistency within the $t$-matrix approaches considered in this paper.
A judicious optimization in achieving convergence is, in fact, especially required for those approaches that implement partial or total self-consistency in the particle-particle propagator $\Gamma$, 
to the extent that these approaches become intrinsically unstable when the temperature approaches $T_{c}$ when one uses the straightforward iterative procedure sketched in Section~\ref{sec:t-matrices} (cf. Fig.~\ref{Figure-2} therein).
The origin of this problem can be highlighted through the following analytic considerations. 

\vspace{0.05cm}
\begin{center}
{\bf 1. General considerations on the iterative procedure}
\end{center}

For definiteness, we shall consider in detail the fully self-consistent $(GG)G$ approach. 
The equations (\ref{equation-G})-(\ref{equation-Rpp_Q}), that need to be self-consistently solved, can be written in a compact way as a functional equation for the self-energy $\Sigma$, in the form:
\begin{equation}
\Sigma(k)=\mathcal{F} \big[ \Sigma(p)\big]\!(k) \, .
\label{equation-SC_functional_eq}
\end{equation}
This is because, once $\Sigma$ is know, $G$ and $\Gamma$ in Eqs.~(\ref{equation-G})-(\ref{equation-Rpp_Q}) can also be readily obtained.
Let us then see what happens when trying to solve Eq.~(\ref{equation-SC_functional_eq}) iteratively. 

Suppose that $\Sigma^{\mathrm{SC}}(k) = \mathcal{F}\big[ \Sigma^{\mathrm{SC}}(p) \big]\!(k)$ is the self-consistent ($\mathrm{SC}$) solution to Eq.~(\ref{equation-SC_functional_eq}) which the iterative method 
is expected to reach.
At a generic iteration step $\mathrm{(i)}$ toward self-consistency, the self-energy $\Sigma^{\mathrm{(i)}}(k)$ deviates from $\Sigma^{\mathrm{SC}}(k)$ by some amount $\delta\Sigma^{\mathrm{(i)}}(k)$:
\begin{equation}
\Sigma^{\mathrm{(i)}}(k) = \Sigma^{\mathrm{SC}}(k) + \delta \Sigma^{\mathrm{(i)}}(k).
\end{equation}
Provided one is close enough to the self-consistent solution, Eq.~(\ref{equation-SC_functional_eq}) can be linearized about $\Sigma^{\mathrm{SC}}$, to write:
\begin{eqnarray}
&& \Sigma^{\mathrm{SC}}(k)+ \delta \Sigma^{\mathrm{(i)}} (k) = \mathcal{F} \big[\Sigma^{\mathrm{(i-1)}}(p) \big]\!(k) 
\nonumber \\
& \simeq & \Sigma^{\mathrm{SC}}(k) + \int dp \, \Bigg[\frac{\delta \mathcal{F}[\Sigma(p)]\!(k)}{\delta \Sigma(p)} \Bigg]_\text{SC} \delta \Sigma^{\mathrm{(i-1)}}(p) \, .
\label{Sigma-expanded}
\end{eqnarray}
Here, the subscript $\mathrm{SC}$ indicates that the functional derivative is taken at self-consistency, and the integral over $p$ contains both an integral over the wave vector $\mathbf{p}$ and a sum over 
the Matsubara frequency $\omega_{n}$. 
This provides a relation between the distance $\delta \Sigma$ from the self-consistent solution between the steps $\mathrm{(i-1)}$ and $\mathrm{(i)}$:
\begin{equation}
\delta \Sigma^{\mathrm{(i)}} (k) = \int dp \, \Bigg[\frac{\delta \mathcal{F}[\Sigma(p)]\!(k)}{\delta \Sigma(p)} \Bigg]_{\mathrm{SC}} \delta \Sigma^{\mathrm{(i-1)}}(p) .
\label{equation-delta_Sigma_i}
\end{equation}
When $\Sigma(k)$ is calculated on a $k$-grid of points, like it is done in our numerical calculations, $\delta \Sigma$ can be regarded as a vector which is acted upon by the functional derivative matrix
$[\delta \mathcal{F}/\delta \Sigma]_{\mathrm{SC}}$. 
The convergence of the iterative procedure, from $\mathrm{(i-1)}$ to $\mathrm{(i)}$ and so on, is then governed by the behavior of this functional derivative matrix. 

To find an explicit expression for $[\delta \mathcal{F}/\delta \Sigma]_{\mathrm{SC}}$, we rewrite it in the form:
\begin{equation}
\frac{\delta \mathcal{F}[\Sigma(p)]\!(k)}{\delta \Sigma(p)} = \frac{\delta \mathcal{F}[\Sigma(p)]\!(k)}{\delta G(p)} \, \frac{\delta G(p)}{\delta \Sigma(p)} \, .
\label{equation-dF/dSigma1}
\end{equation}
Here, the factor on the right is given by
\begin{equation}
\frac{\delta G(p)}{\delta \Sigma(p)} = G(p)^{2} 
\end{equation}
with the use of Eq.~(\ref{equation-G}),  
while the factor on the left can be calculated by recalling that (cf. Eqs.~(\ref{equation-Sigma_k}) and (\ref{equation-SC_functional_eq}))
\begin{equation}
\mathcal{F} \big[ \Sigma(p) \big](k) = \Sigma(k) = - \! \int \!\! dQ  \,\Gamma(Q) G(Q-k) \, ,
\end{equation}
yielding
\begin{equation}
\frac{\delta \mathcal{F}[\Sigma(p)]\!(k)}{\delta G(p)} = - \Gamma(p+k) - \! \int \! dQ \, \frac{\delta \Gamma(Q)}{\delta G(p)} G(Q-k) .
\label{equation-dF/dSigma2}
\end{equation}
In this expression, the functional derivative of the particle-particle propagator $\Gamma(Q)$ can be obtained from Eq.~(\ref{equation-Gamma_Q})
\begin{equation}
\frac{\delta \Gamma(Q)}{\delta G(p)} = - \Gamma(Q)^2 \, \frac{\delta R_{\mathrm{pp}}(Q)}{\delta G(p)} ,
\label{equation-dGamma/dG}
\end{equation}
where the functional derivative of the regularized particle-particle bubble $R_{\mathrm{pp}}(Q)$ is obtained from Eq.~(\ref{equation-Rpp_Q})
\begin{equation}
\frac{\delta R_{\mathrm{pp}}(Q)}{\delta G(p)} = 2 \,G(Q-p ) .
\label{delta-R-delta-G}
\end{equation}
Grouping all the above results together, Eq.~(\ref{equation-dF/dSigma1}) becomes eventually: 
\begin{eqnarray}
\frac{\delta \mathcal{F}[\Sigma(p)]\!(k)}{\delta \Sigma(p)} & = & - \, G(p)^{2} \bigg[ \Gamma(p+k) 
\label{equation-dF/dSigma1-explicit} \\
&- & 2 \int \!\! dQ  \,\Gamma(Q)^{2} G(Q-k) G(Q-p) \bigg] .
\nonumber 
\end{eqnarray}

When this result is used in Eq.~(\ref{equation-delta_Sigma_i}), $\delta \Sigma^{(i)}$ therein can be conveniently split in two terms
\begin{equation}
\delta \Sigma^{\mathrm{(i)}} (k) = \delta \Sigma_{1}^{\mathrm{(i)}} (k) + \delta \Sigma_{2}^{\mathrm{(i)}} (k) ,
\end{equation}
where we have defined
\begin{eqnarray}
\delta \Sigma_{1}^{\mathrm{(i)}} (k) & = & - \int \!\! dp \, G(p)^{2} \Gamma(p+k) \, \delta \Sigma^{\mathrm{(i-1)}} (p)
\label{equation-delta_Sigma1} \\
\delta \Sigma_{2}^{\mathrm{(i)}} (k) & = & 2 \int \!\! dp \, G(p)^{2} \, \int \!\! dQ \, \Gamma(Q)^{2}
\nonumber \\
& \times & G(Q-k) G(Q-p) \, \delta \Sigma^{\mathrm{(i-1)}} (p) .
\label{equation-delta_Sigma2}
\end{eqnarray}  

Suppose now that $T=T_{c}$. 
The Thouless criterion (\ref{equation-Thouless_criterion}) then implies that the particle-particle propagator $\Gamma(Q)$ has a pole for $Q=0$, such that one expects $\Gamma(Q)$ to remain strongly peaked in the vicinity of $Q = 0$. 
The expressions (\ref{equation-delta_Sigma1}) and (\ref{equation-delta_Sigma2}) can be simplified accordingly, by setting to zero the arguments of the particle-particle propagators in the smooth functions that multiply them.
For the term (\ref{equation-delta_Sigma1}) we thus have:
\begin{eqnarray}
\delta \Sigma_1^{\mathrm{\mathrm{(i)}}} (k) & \simeq & - G(-k)^2 \delta \Sigma^{\mathrm{\mathrm{(i-1)}}}(-k) \int \! dp \, \Gamma(p+k)
\nonumber \\
& = & - C \,\, G(-k)^2 \, \delta \Sigma^{\mathrm{(i-1)}}(-k)
\end{eqnarray}
where $C$ is the Tan's contact according to Eq.~(\ref{equation-contact}). 
This term poses no problem to the convergence, since the quantity that multiplies $\delta \Sigma^{\mathrm{(i-1)}}(-k)$ is finite. 
For the term (\ref{equation-delta_Sigma2}) we instead obtain: 
\begin{eqnarray}
\delta \Sigma_2^{\mathrm{(i)}} (k) & \simeq & 2 G(-k) \, \int \!\! dQ \, \Gamma(Q)^{2} 
\nonumber \\
& \times & \int \!\!dp \, G(p)^{2} G(-p) \, \delta \Sigma^{\mathrm{(i-1)}}(p) 
\label{equation-delta_Sigma2_Tc}
\end{eqnarray}
where the factor
\begin{equation}
\int \!\! dQ \, \Gamma(Q)^{2} = \int \frac{d\mathbf{Q}}{(2\pi)^3} T \sum_\nu \Gamma(\mathbf{Q},\Omega_{\nu})^{2}
\label{equation-int_Gamma2}
\end{equation}
is infrared divergent at $T=T_{c}$ even in three dimensions.
This is because, for the term with $\Omega_{\nu}=0$ therein, $\Gamma(\mathbf{Q},\Omega_\nu=0)\sim \mathbf{Q}^{-2}$ when $\mathbf{Q} \rightarrow 0$ at $T=T_c$ \cite{PPSC-2002}. 

This divergence represents a problem for the convergence of the iterative algorithm, because it implies that, no matter how close the step $\mathrm{(i-1)}$ might be to the self-consistent solution, 
the step $\mathrm{(i)}$ is bound to run infinitely away from it. 
This problem can affect the convergence of the iterative algorithm also for temperatures $T \gtrsim T_{c}$, depending on how much $\Gamma(Q)$ is peaked about $Q=0$. 

The only (partially) self-consistent $t$-matrix approach not affected by this problem is the $(G_{0} G_{0})G$ approach. 
This is because in this approach, by construction, the particle-particle propagator $\Gamma$ coincides with the bare $\Gamma_{0}$, whereby $\delta \Gamma_{0}/\delta G = 0$.
As a consequence, the second term on the right-hand side of Eq.~(\ref{equation-dF/dSigma2}) identically vanishes and with it the divergent factor in Eq.~(\ref{equation-delta_Sigma2_Tc}).

We pass now to show how this problem can be overcome in practice, both for $T>T_{c}$ and $T \rightarrow T_{c}$.

\vspace{0.05cm}
\begin{center}
{\bf 2. Improved method for $T>T_{c}$}
\end{center}

Although the iterative approach (\ref{equation-delta_Sigma_i}) to solve Eq.~(\ref{equation-SC_functional_eq}) cannot converge at exactly $T=T_{c}$, there exists a simple method to make it converge for $T \gtrsim T_{c}$, 
with the factor (\ref{equation-int_Gamma2}) keeping a finite (albeit large) value. 
This method, which has already been used for the self-consistent calculations of electronic structures in atoms and molecules \cite{Ferreira-1980}, consists in redefining the iterative steps in terms of the weighted sum:
\begin{equation}
\Sigma^{\mathrm{(i)}}(k) = \alpha \, \mathcal{F} \big[ \Sigma^{\mathrm{(i-1)}}(p)\big]\!(k)+(1-\alpha) \Sigma^{\mathrm{(i-1)}}(k)
\label{equation-weighted_sum}
\end{equation}
where the weight factor $\alpha$ ranges between $0$ and $1$. 
This method is found to reduce the effects of the divergence due to the term (\ref{equation-delta_Sigma2_Tc}), thereby making the iterative process to converge even sufficiently close to $T_c$.
Nevertheless, the method fails upon approaching $T_c$, because smaller and smaller values of $\alpha$ are needed for attaining convergence.
A smaller value of $\alpha$, in turn, implies that more iterative steps are required for convergence, since the contribution of a single step, too, becomes smaller and smaller. 
In practice, we have found that this method can conveniently been used down to temperatures for which \mbox{$(T-T_c)/T_c$} is of order $1\%$, before it becomes numerically too demanding.

\vspace{0.05cm}
\begin{center}
{\bf 3. Improved method for $T=T_{c}$}
\end{center}

Alternatively, exactly at $T=T_c$ one can rely on a different method that avoids the convergence problem discussed above. 
This method can be summarized as follows:

\noindent 
(i) One begins by fixing the value $n$ of the density, in terms of which one obtains the Fermi wave vector $k_{F}=(3 \pi^{2} n)^{1/3}$ and the Fermi energy $E_{F}=k_{F}^{2}/(2m)$.
One also fixes a guess value $T_{\mathrm{g}}/T_{F}$ for the temperature in units of the Fermi energy, as well of the ratio $\mu/T_{\mathrm{g}}$ between the chemical potential 
and the guess temperature $T_{\mathrm{g}}$ (recall that we have set the Boltzmann constant $k_{B}$ equal to unity throughout).

\noindent 
(ii) Next, one replaces the particle-particle propagator of Eq.~(\ref{equation-Gamma_Q}) with the following expression
\begin{equation}
\tilde{\Gamma}(Q)^{-1} = R_{\mathrm{pp}} (Q)-R_{\mathrm{pp}}(Q=0) \, ,
\label{equation-Gamma_new}
\end{equation}
in such a way that $\tilde{\Gamma}(Q=0)^{-1}=0$ by construction.
This implies that the Thouless criterion (\ref{equation-Thouless_criterion}) is always satisfied by the modified particle-particle propagator (\ref{equation-Gamma_new}), \emph{no matter} what was the initial
guess for $T_{\mathrm{g}}/E_{F}$.  

\noindent 
(iii) At this point one can proceed and perform the iterative procedure toward self-consistency on the set of equations (\ref{equation-G})-(\ref{equation-Rpp_Q}), with $\Gamma(Q)$ therein replaced by $\tilde{\Gamma}(Q)$.
It will be shown below that this replacement avoids the occurrence of the infrared divergence that plagues instead the expression (\ref{equation-delta_Sigma2_Tc}) obtained in terms of the original $\Gamma(Q)$.
 
\noindent 
(iv) Once self-consistency has been achieved with this modified set of equations, one can obtain the modified density as $\tilde{n}= - 2 \, \tilde{G}(\mathbf{r}=0,\beta^{-})$, 
and the modified scattering length $\tilde{a}_{F}$ from the expression
\begin{equation}
\frac{1}{\tilde{a}_{F}} = - \, \frac{4 \pi}{m} \, \tilde{R}_{\mathrm{pp}}(Q=0) 
\label{equation-coupling_new}
\end{equation}
where $\tilde{R}_{\mathrm{pp}}$ is obtained from Eq.~(\ref{equation-Rpp_Q}) with $\tilde{G}$ replacing $G$.
The result (\ref{equation-coupling_new}) follows directly from the Thouless criterion corresponding to the modified density $\tilde{n}$.

\noindent 
(v) Finally, in terms of $\tilde{n}$ one obtains the modified Fermi wave vector $\tilde{k}_{F}=(3 \pi^{2} \tilde{n})^{1/3}$ and the modified Fermi energy $\tilde{E}_{F}=\tilde{k}_{F}^{2}/(2m)$.
The desired value of the critical temperature is then obtained by
\begin{equation}
\frac{T_{c}}{E_{F}} = \frac{T_{\mathrm{g}}}{E_{F}} \,\, \frac{E_{F}}{\tilde{E}_{F}}
\label{desired-critical-temperature}
\end{equation}
in terms of the initial guess $T_{\mathrm{g}}/E_{F}$.
From Eq.~(\ref{equation-coupling_new}) the corresponding coupling value is given by:
\begin{equation}
\frac{1}{k_{F} a_{F}} = \frac{1}{\tilde{k}_{F} \tilde{a}_{F}} =  \frac{1}{k_{F} \tilde{a}_{F}} \, \frac{k_{F}}{\tilde{k}_{F}} = - \frac{4 \pi \tilde{R}_{\mathrm{pp}}(Q=0)}{m k_{F}} \, \frac{k_{F}}{\tilde{k}_{F}} \, .
\label{corresponding-coupling}
\end{equation}

There remains to explain the reason why this method is not plagued by the convergence problem discussed above for the iterative procedure.
The point is that, using the modified particle-particle propagator (\ref{equation-Gamma_new}) in the place of the original one, one also modifies the structure of the functional derivative in Eq.~(\ref{equation-delta_Sigma_i}). 
Specifically, for the functional derivative of $\tilde{\Gamma}(Q)$ with respect to $\tilde{G}(p)$ one obtains (cf. Eq.~(\ref{equation-dGamma/dG})):
\begin{eqnarray}
\frac{\delta \tilde{\Gamma}(Q)}{\delta \tilde{G}(p)} & = & - \tilde{\Gamma}(Q)^{2} \, \frac{\delta \big(\tilde{R}_{\mathrm{pp}}(Q) - \tilde{R}_{\mathrm{pp}}(Q=0)\big)}{\delta \tilde{G}(p)} 
\nonumber \\
& = & - 2 \, \tilde{\Gamma}(Q)^{2} \, \big[\tilde{G}(Q-p)-\tilde{G}(-p) \big] \, .
\label{equation-dGamma_new/dG}
\end{eqnarray}
The corresponding variation of the self-energy related to this functional derivative then becomes (cf. Eq.~(\ref{equation-delta_Sigma2})):
\begin{eqnarray}
\delta \tilde{\Sigma}_{2}^{\mathrm{(i)}} (k) & = & 2 \int \!\! dp \, \tilde{G}(p)^{2}  \int \!\! dQ \, \tilde{\Gamma}(Q)^{2}  
\label{final-equation} \\
& \times & \tilde{G}(Q-k) \big[\tilde{G}(Q-p)-\tilde{G}(-p) \big] \, \delta \tilde{\Sigma}^{\mathrm{(i-1)}} (p) .
\nonumber 
\end{eqnarray}
Comparing this result with Eq.~(\ref{equation-delta_Sigma2}), one notes that the singular behavior of $\tilde{\Gamma}(Q)^{2}$ for $Q \rightarrow 0$ is now suppressed by the presence of the factor 
$[\tilde{G}(Q-p)-\tilde{G}(-p)]$. 
This feature makes it possible to reach convergence \emph{exactly at} $T=T_{c}$ with a limited number of iterations, without the need for the weighted sum of Eq.~(\ref{equation-weighted_sum}).

\section{COMPARISON WITH THE PSEUDO-GAP APPROXIMATION AT $T_c$}
\label{sec:appendix-C}
\vspace{-0.2cm}

The $(GG_{0})G_{0}$ and $(GG)G_{0}$ $t$-matrix approaches have almost invariably been implemented in the literature using a set of approximations (sometimes referred to as the ``pseudo-gap approximation"), which considerably simplify the numerical calculations.  
Specifically, close to $T_{c}$ where the particle-particle propagator $\Gamma(Q)$ is strongly peaked about $Q=0$, the fermionic self-energy (\ref{equation-Sigma_k}) has been approximated as follows \cite{Maly-1999,Chen-2005}:
\begin{small}
\begin{eqnarray}
\Sigma(\mathbf{k},\omega_{n}) & = &-\int\!\!\frac{d\mathbf{Q}}{(2\pi)^3} T \sum_\nu \Gamma(\mathbf{Q},\Omega_{\nu}) \, G_{0}(\mathbf{Q-k},\Omega_{\nu}-\omega_{n})\label{self} 
\nonumber \\
& \approx & G_{0}(-\mathbf{k},-\omega_{n}) \! \left( \! - \! \int\!\!\frac{d\mathbf{Q}}{(2\pi)^3} T \sum_\nu \Gamma(\mathbf{Q},\Omega_{\nu}) e^{i \Omega_{\nu} 0^+}\right)
\nonumber \\
& \equiv & G_{0}(-\mathbf{k},-\omega_{n})  \, \Delta^{2}_{{\rm pg}} \, .
\label{pgap}
\end{eqnarray}
\end{small}
Due to this approximation for the self-energy, the dressed single-particle propagator $G$ (and thus also the equation for the particle number) coincides with that of BCS theory, with the pseudo-gap energy $\Delta_{\rm pg}$ now playing the role in the normal phase of the BCS gap $\Delta$ in the superfluid phase.  
In addition, the Thouless criterion (\ref{equation-Thouless_criterion}) for the dressed $\Gamma$ coincides with the BCS gap equation (again with the replacement $\Delta \rightarrow \Delta_{\rm pg}$) for the $(GG_{0})G_{0}$ approach \cite{Maly-1999,Chen-2005}, or with the BCS gap equation plus additional corrections (which become anyway negligible both in the BCS and BEC limits) for the $(GG)G_{0}$ approach \cite{Micnas-2014}. 
A further approximation, which is usually adopted within the pseudo-gap approximation when calculating $\Delta_{\rm pg}$ by means of Eq.~(\ref{pgap}), is the use of 
an expansion of $\Gamma({\bf Q},\Omega_{\nu})$ for small values of ${\bf Q}$ and $\Omega_{\nu}$. 

\begin{figure}[t]
\begin{center}
\includegraphics[width=6.5cm,angle=0]{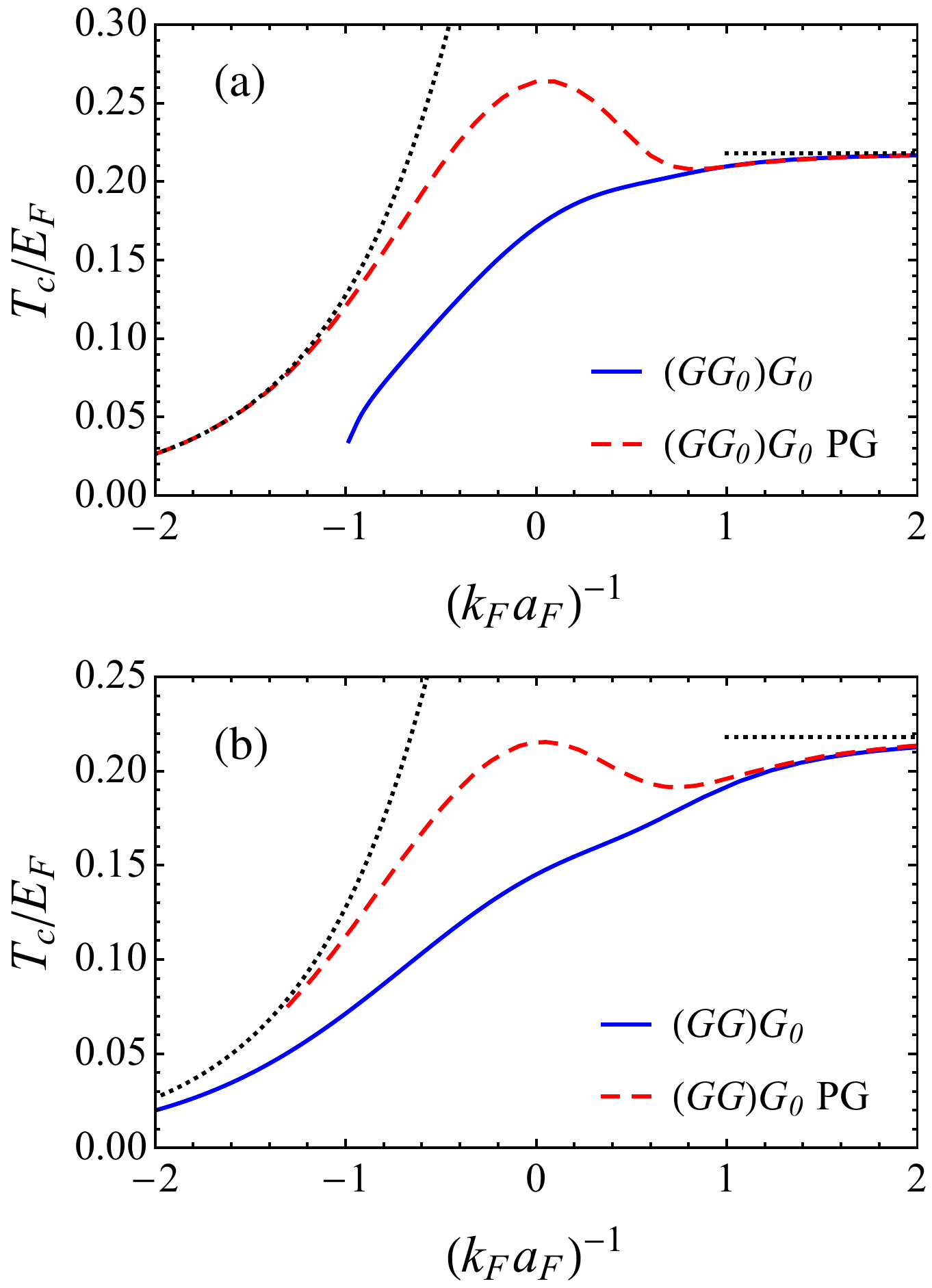}
\caption{(Color online) Critical temperature $T_{c}$ vs the coupling $(k_{F} a_{F})^{-1}$ for the $(GG_{0})G_{0}$ (panel a) and $(GG)G_{0}$ (panel b) approaches (full lines), 
                                    and for their corresponding pseudo-gap (PG) approximation (dashed lines). 
                                    The BCS and BEC critical temperatures are also reported for comparison (dotted curves on the left and right sides, respectively).
                                    The data for the pseudo-gap approximation to the $(GG_{0})G_{0}$ and $(GG)G_{0}$ approaches are taken from Refs.~\cite{Levin-2010} and \cite{Micnas-2014}, respectively.}
\label{Figure-13}
\end{center}
\end{figure} 

It should, however, be remarked that the pseudo-gap approximation (\ref{pgap}) appears justified only in the strong-coupling (BEC) regime $(k_{F} a_{F})^{-1} \gtrsim +1$, where the large fermionic energy scale  $|\mu|$ (with $\mu <0$) dominates over the bosonic energy scales and the approximation (\ref{pgap}) becomes fully correct.
This can also be verified numerically as shown in Fig.~\ref{Figure-13}, where a comparison is presented for the calculation of $T_{c}$ between the complete $(GG_{0})G_{0}$ and $(GG)G_{0}$ approaches (full lines) and their corresponding pseudo-gap approximations (dashed lines).
This comparison shows that a good agreement between the complete and approximate calculations occurs only for $(k_{F} a_{F})^{-1} \gtrsim 1$, while significant deviations result both at intermediate and weak couplings. 
 
From Fig.~\ref{Figure-13} one also notices that, in the weak-coupling (BCS) regime $(k_{F} a_{F})^{-1} \lesssim -1$, the curves for $T_{c}$ obtained within the pseudo-gap approximation converge rapidly to the corresponding BCS curve for $T_{c}$. 
This is because $\Delta_{\rm pg}(T)$ is bounded by the value $\Delta_{0}$ of the BCS gap at $T=0$ \cite{Chen-2005}, which in turn vanishes exponentially in the weak-coupling limit. 
This implies that the approximate self-energy (\ref{pgap}), too, vanishes exponentially, in such a way that the BCS result for $T_{c}$ is recovered. 
Without the use of the approximation (\ref{pgap}), in weak coupling the self-energy would instead approach the value $\Sigma\simeq 2\pi a_{F} n/m$ associated with a mean-field shift. 

For the complete $(GG_{0})G_{0}$ approach, whereby this shift appears in just one of the two single-particle propagators that enter the particle-particle bubble in $\Gamma$, the equation for $T_{c}$ corresponds to that of a Fermi system in the presence of a chemical potential imbalance $\delta \mu$.
This equation is known to have no solution when this imbalance about exceeds the BCS gap $\Delta_{0}$ of the balanced system at $T=0$ \cite{Sarma-1963}.  
Given the exponential dependence of $\Delta_{0}$ on coupling, to be contrasted with the linear dependence $\delta \mu \simeq 2\pi a_{F} n/m$ associated with the mean-field shift, the condition $\delta \mu > \Delta_{0}$
is readily met in the weak-coupling regime.
This explains why, in the weak-coupling regime, the $(GG_{0})G_{0}$ approach does not admit solution for $T_{c}$, as it was already noted in the discussion of Fig.~\ref{Figure-3}. 
For the complete $(GG)G_{0}$ approach, on the other hand, the differences with respect to its pseudo-gap approximation are overall less pronounced, albeit still significant. 




\begin{thebibliography}{99}
\bibitem{footnote-2} Additional care is required in weak coupling, where the expressions that include the factor $e^{\mu \tau}$ become large when $\tau \simeq \beta$ and may then 
                                lead to numerical errors. 
                                In this case, for the $\mathbf{k} \to \mathbf{r}$ Fourier transform it is useful to replace the factor $e^{\mu \tau}$ in Eqs.~(\ref{equation-G0a}) and (\ref{equation-G0a_r}) for
                                $G_{0}^{\mathrm{(a)}}$ by its expansion $(1+\mu \tau + \cdots)$ for $\tau \to 0^{+}$ .
\bibitem{footnote-3} Also in this case, additional care is required to handle in weak coupling the expressions containing the factor $e^{2\mu \tau}$. 
                                In this case, the separation of $\Gamma_\mathrm{sc}(\mathbf{Q},\tau)$ in numeric and analytic parts  like in Eq.~(\ref{equation-Gamma'_sc_Q_tau}) has to be performed 
                                only for $\tau \simeq 0$, while $\Gamma_\text{sc}(\mathbf{Q},\tau)$ has to be calculated as a whole for larger values of $\tau$. 
                                Moreover, when performing the $\mathbf{Q} \to \mathbf{r}$ Fourier transform, it is again useful to replace the factor $e^{2\mu \tau}$ in Eqs.~(\ref{equation-Gamma0a_Q_tau}) and 
                                (\ref{equation-Gamma0a_r_tau}) for $\Gamma_{\mathrm{sc}}^{\mathrm{(a)}}$ with its expansion $(1+2\mu \tau + \cdots)$ for $\tau \to 0^{+}$.
\end{thebibliography}
\end{document}